\documentclass[draftcls,  onecolumn, 11 pt]{IEEEtran}
\usepackage{amsfonts, cite}
\usepackage{amssymb}
\usepackage{color}
\usepackage{subfig}
\usepackage{scalefnt}
\usepackage{amsmath}
\usepackage{amsfonts}
\usepackage{colordvi}
\usepackage{verbatim}
\usepackage{graphicx}
\usepackage{latexsym}
\usepackage{bm}
\usepackage{cite}

\usepackage{times}

\usepackage{algpseudocode}

\newtheorem{theorem}{Theorem}
\newtheorem{lemma}{Lemma}

\newtheorem{corollary}{Corollary}

\title{Power Control and Interference Management in Dense Wireless Networks}%
\author{Ehsan~Karamad,~\IEEEmembership{Student~Member,~IEEE,
} Raviraj~S.~Adve,~\IEEEmembership{Senior~Member,~IEEE,}~and~Jerry~Chow%
\thanks{E. Karamad and R. S. Adve are with Edwards S. Rogers Sr.~Dept. of Electrical and Computer Engineering, University of Toronto email: \{ekaramad, rsadve\}@comm.utoronto.ca}%
\thanks{J. Chow is with BLiNQ Wireless Inc, Plano, TX, USA email: jerry.chow@blinqnetworks.com}}
\IEEEaftertitletext{\vspace{-1.25\baselineskip}}
\begin{document}
\maketitle \thispagestyle{empty}

\begin{abstract}
We address the problem of interference management
and power control in terms of maximization of a general utility function.
For the utility functions under consideration, we
propose a power control algorithm based on a fixed-point
iteration; further, we prove local convergence of the algorithm in the
neighborhood of
the optimal power vector. Our algorithm has several benefits
over the previously studied works in the literature: first, the
algorithm can be applied to problems other than network utility
maximization (NUM), e.g., power control in a relay network;
second, for a network with $N$ wireless transmitters, the
computational complexity of the proposed algorithm is $\mathcal{O}(N^2)$
calculations per iteration (significantly smaller than the $\mathcal{O}(N^3)
$ calculations
for Newton's iterations or gradient descent approaches). Furthermore, the
algorithm converges very fast (usually in less than 15 iterations), and in
particular, if initialized close
to the optimal solution, the convergence speed is much faster.
This suggests the potential of tracking variations in slowly
fading channels. Finally, when implemented in a distributed
fashion, the algorithm attains the optimal power vector with a
signaling/computational complexity of only $\mathcal{O}(N)$ at each node.

\end{abstract}

\section{Introduction}
\label{sec:introduction}
Interference control is one of the major problems of today's
dense wireless networks; indeed, the latest releases of the Long Term
Evolution (LTE) and 802.16 (WiMax) standards take measures
to control the interference environment in dense network
implementations~\cite{LTE_WIMAX:2010,IEEE802_16mSDD}. Intra-cell interference, caused by other cellular
users within the cell,
occurs in cellular networks with non-orthogonal channel assignment within a
single cell, e.g., code division multiple access (CDMA) networks. In more
recent cellular network standards, this problem is resolved through
orthogonal channel access assignments, e.g., orthogonal frequency division
multiple access (OFDMA)~\cite{LTE_WIMAX:2010,IEEE802_16mSDD}. Inter-cell interference, i.e., the interference received from adjacent cells, is a significant problem even in modern cellular networks with intra-cell interference management.  Essentially, the problem of designing practical and efficient interference control techniques has become extremely important in the design of next generation wireless cellular networks.

\subsection{Previous Work}
The power control problem to reduce co-channel interference was proposed as a linear programming problem as early as 1964~\cite{Brock:1964}, wherein the authors considered solving for the minimum transmission power given a required signal to noise and interference ratio (SINR). In the years that followed, researchers investigated power control in different applications. Most of the work up to the early 1990s focused on techniques to maintain the received power at a constant level~\cite{Sakamoto:1988} (e.g., to tackle the near-far problem in CDMA networks).
The work of Zander~\cite{Zander:1992} and, soon after, the distributed power control (DPC) algorithm by Foschini~\emph{et al.}~\cite{Foschini:1993} were early solutions for SINR assignment in cellular networks. These preliminary works consider the problem of setting equal~\cite{Zander:1992} and arbitrary~\cite{Foschini:1993} target SINRs for wireless users in cellular networks. Although these solutions efficiently solve the power control problem in distributed cellular networks, they do not consider if the assumed initial SINR target is feasible. Moreover, the assumption of target SINR levels is mainly applicable to voice networks wherein a minimum SINR is required to maintain communications. In recent cellular network applications, more sophisticated power assignment techniques are necessary to ensure adequate quality of service (QoS) throughout the network.

One body of research work following~\cite{Zander:1992} and~\cite{Foschini:1993} emphasize the distributed aspects of the power control algorithm at the cost of optimality of the solution (e.g., see ~\cite{Xiao:2003, Saraydar:2001, Goodman:2000} and the reference therein). For instance, the work in~\cite{Xiao:2003} proposes a variant of DPC and proves its convergence. In their algorithm, the wireless nodes update the transmit powers based on the observed interference such that a~\emph{node utility} defined as a concave function of SINR minus a linear power cost, is maximized, i.e.,  distributed maximization of the Lagrangian for an optimization problem where, for arbitrary power costs (dual variables), the obtained transmit powers are not necessarily optimal. Other works, such as~\cite{Saraydar:2001}, investigate game theoretic approaches in power control converging to a Nash equilibrium. The analysis of convergence rates of some distributed power control techniques based on fixed-point iterations can be found in~\cite{Johansson:2012} and the references therein.

A second category of research works investigate globally optimal power control in interference networks, especially the works of Chiang~\emph{et al.}~\cite{MChiang:2004,MChiang:2006,MChiang:2008,MChiangBook}. The work in~\cite{MChiangBook} covers some of the major problems and the corresponding solutions. The research works in this category attempt to characterize the set of feasible transmit powers as a convex set and propose a convex optimization framework to  find the optimal power levels for the wireless transmitters. One of the standard methods of describing the feasible set of transmit powers is to consider the spectral radius of a matrix made of network-wide parameters~\cite{Zander:1992}. Chiang~\emph{et al.} use this notion to propose a distributed power control algorithm in~\cite{MChiang:2006,MChiang:2008,MChiang:2004}. Their work follows a network utility maximization (NUM)~\cite{Kelly:1998} framework where they use gradient ascend methods for a class of concave utility functions. While such algorithms are proven to converge to the globally optimal solution, the convergence speed is rather slow due to the reliance on the updates on the gradient of the Lagrangian. In addition, the algorithm involves finding the~\emph{Perron} root~\cite{Graham:1987} of a matrix at each iteration. For an $N \times N$ matrix, finding the root is known to be of order $\mathcal{O}(N^3)$ in complexity and therefore, not properly suited to larger network sizes.

A different approach in solving network utility maximization problems is to formulate an equivalent problem whose feasible set is convex in its variables. For instance, it has been shown that the set of feasible power vectors is $\log$-convex~\cite{Boche:2004, MChiangBook}, or, for a variety of utility functions, the optimization problem is convex if the problem is formulated in the inverse of power variables~\cite{MChiangBook}.  Using the former technique, the work in~\cite{Boyd:2003} proposes a greedy power control technique based on the direction of the gradient of the objective function. As mentioned before, the major drawback of gradient-based methods is the potentially slow rate of convergence to the optimal solution.

More recent works on power control focus on fast converging algorithms and non-convex cases not addressed before. For instance, the work in~\cite{Qian:2009} proposes the MAPEL algorithm which solves the weighted sum-rate maximization problem. They express the objective function in terms of a product of exponentiated linear fraction functions; this leads to  generalized linear fraction function programming (GLFP). However, there is, in general, a trade-off between the speed of convergence and global optimality in the solution~\cite{Qian:2009}. Another work in~\cite{Dahrouj:2011} addresses the power control problem for a weighted sum-rate objective function. The authors propose iterative function evaluation for the solution and, empirically, show the convergence of their method.

Regarding the speed of convergence in general resource allocation problems, recent works have proposed practical implementations of Newton's methods to attain the optimal solution with as little computational overhead as possible. Due to its reliance on fixed point iteration, Newton's method enjoys a quadratic convergence rate near the root. The work in~\cite{Boyd:2010} investigates assignment of spectral resources in wireless cellular network where the authors improve the implementation of the Newton's method for the barrier function approach in solving convex optimization problems. In particular, for the system model under consideration, the inverse of the Jacobian matrix can be calculated efficiently. Another work based on the same approach has been presented in~\cite{Wang:2012} where, similar to~\cite{Boyd:2010}, efficient calculation of the Newton's iterations leads to fast converging and numerically efficient resource allocators for OFDM-based cognitive radio networks.

\subsection{Our Contributions}
Motivated by recent works on fast resource allocation in cellular wireless networks, in this paper, we will develop a fast-converging power control algorithm. We start by formulating a utility maximization problem (containing NUM problems as a subclass) and argue that for a class of utility functions concave in the logarithm of SINR whose logarithm of gradient in terms of $\log$-SINR is Lipschitz continuous~\cite{Searcoid:2006}, the problem can be solved efficiently (this includes most commonly used utility functions). We then propose an iterative algorithm based on fixed-point iterations, for problems with and without a minimum/maximum power constraint at the wireless transmitters. We prove that our proposed algorithm converges to the optimal power vector and the convergence rate is at least~\emph{linear}\footnote{That is, at iteration $k$, the norm of error scales as $\mathcal{O}(a^k)$ where $0 < a < 1$}. Moreover, for a network with $N$ wireless transmitters, the complexity of our solution is $\mathcal{O}(N^2)$. This is more efficient and scalable than gradient-based methods. In summary, our contributions are:
\begin{itemize}
\item{Formulation of the power control problem in an interference network for a~\emph{general} concave utility function.}
\item{Providing necessary and sufficient conditions for the optimality of the transmit power vector.}
\item{Presenting a fast and low-complexity algorithm which converges to the optimal solution for problems with or without power constraints.}
\item{Proof of convergence for the power control algorithm with or without power constraints.}
\end{itemize}
As will be illustrated in the numerical results section, one key benefit in using the proposed iterative power control is its capability in tracking slowly fading wireless channels. Moreover, since our formulation of the power control problem is based on a general utility function (as opposed to sum-utility), power control for more sophisticated network structures is possible, e.g., power control in cooperative cellular networks.

\subsection{Notation}
We use lowercase and uppercase bold letters to present column vectors and matrices, respectively. In line with this, $(\mathbf{x})_i$ (or simply $x_i$) represents the $i$th element of the column vector $\mathbf{x}$ and $(\mathbf{A})_{i,j}$ (or simply $A_{i,j}$) is the element at row $i$ and column $j$ of matrix $\mathbf{A}$.
Furthermore,  $\mathbf{x}^T$ and $\mathbf{A}^T$ are transposes of the vector $\mathbf{x}$ and matrix $\mathbf{A}$, respectively. We use the function $\mathbf{D}(\mathbf{x}) : \mathbb{R}^N \rightarrow \mathbb{R}^{N \times N}$ to define the diagonal matrix with the elements of vector $\mathbf{x}$ as its diagonal elements, i.e., $\left(\mathbf{D}(\mathbf{x})\right)_{i,i} = (\mathbf{x})_i$.

We extend the basic operations in field of real numbers $\mathbb{R}$ to the space $\mathbb{R}^N$ as follows: For any two vectors $\mathbf{x}$ and $\mathbf{y}$ in $\mathbb{R}^N$, we define the result of $\mathbf{x}\mathbf{y}$ as a vector in $\mathbb{R}^N$ whose $i$th element equals $(\mathbf{x})_i (\mathbf{y})_i$, i.e., the operation represents an element-wise multiplication of the two vectors. Similarly, we extend scalar functions of real numbers to the vectors in an element-wise fashion. For instance, $\left(e^{\mathbf{x}} \right)_i = e^{(\mathbf{x})_i}$, and $(\mathbf{x}^{k})_i = (\mathbf{x})_i^{k}$ for $k \in \mathbb{R}$.

For any scalar function $f(x)$ with $x, f(x) \in \mathbb{R}$, $f \in C^2$ \footnote{ $C^2$ indicates the class of functions with continuous second derivative.}, $f'(x)$ and $f''(x)$ refer to the first and second derivatives at $x$. For a scalar function $f(\mathbf{x})$ with $\mathbf{x} \in \mathbb{R}^N$ and $f(\mathbf{x}) \in \mathbb{R}$,  $\triangledown_{\mathbf{x}} f \in \mathbb{R}^N$ and $\triangledown_{\mathbf{x}}^2 f \in \mathbb{R}^{N \times N}$ are the gradient vector and Hessian matrix of function $f$ with respect to $\mathbf{x}$. Subsequently, for a vector function $\mathbf{f}(\mathbf{x}): \mathbb{R}^{N} \rightarrow \mathbb{R}^N$, $\mathbf{f}'(\mathbf{x}) \in \mathbb{R}^{N \times N}$ denotes its Jacobian matrix where $\left(\mathbf{f}'(\mathbf{x})\right)_{i,j} = \frac{\partial (\mathbf{f}(\mathbf{x}))_i}{\partial x_j}$.

Finally, for a square and symmetric matrix $\mathbf{A}$, $\mathbf{A} \succeq 0$ and $\mathbf{A} \preceq 0$ imples $\mathbf{A}$ is positive or negative semi-definite, respectively. For a given matrix $\mathbf{A}$ or vector $\mathbf{x}$, $\mathbf{A} \ge 0$ and $\mathbf{x} \ge 0$ means all the elements of the matrix or vector are non-negative.

\subsection{Structure of the Paper}
\label{sec:struct_paper}
The rest of this paper is organized as follow: Section~\ref{sec:system_model} presents our system model and the formulation of the optimal power control problem. It also presents the form of the optimal solution. In Section~\ref{sec:power_control_algorithm}, we propose the power control algorithm achieving the power vector satisfying the optimality conditions presented in Section~\ref{sec:system_model} and provide sufficient conditions on the convergence of the proposed algorithm.  Section~\ref{sec:numerical_results} is dedicated to numerical results based on a practical wireless network model, thereby validating the accuracy and efficiency of our proposed algorithm. Finally, Section~\ref{sec:conclusion} presents a summary of our contributions and concludes this paper.

\section{System Model}
Here we describe the system model under investigation. Then, we propose the utility maximization problem and the corresponding optimality conditions for the solution.

\subsection{Network Model}
\label{sec:system_model}
We investigate a wireless network where $N$ wireless contenders are communicating with their own destination over a shared medium, i.e., a logical network resource which is used by all wireless transmitters. We do not make any assumptions on uplink or downlink transmission or the method of channel access in the network, e.g., frequency/time duplexing or CDMA. Our methodology is based on the channels between each transmitter and all the receivers. We assume that vector $\mathbf{h}$ describes the power of the~\emph{direct channel}, i.e., $h_i$ is the channel magnitude squared for the link between wireless node $i$ to its own destination. The interference channel is described by a matrix $\mathbf{H}$ where $H_{i,j}$ is the (interference) channel magnitude squared from the link from node $j$ to the destination of node $i$. Notice that we allow for self-interference, i.e., $H_{i,i} > 0$~\footnote{As an example, phase noise at the receiver can be modeled as an additive term with a variance proportional to the received signal power. Otherwise, an assumption of $H_{i,i} > 0$ leads to a finite cap on achieved SINR.}.

To simplify the formulation, we define the normalized interference matrix $\mathbf{V} = \mathbf{D}(\mathbf{h})^{-1} \mathbf{H}$. In this regard, the normalized received interference vector $\mathbf{q}$ (with $q_i$ as the normalized interference power received at the destination of node $i$) is defined as
\begin{align}
\label{eq:defn_q}
\mathbf{q} = \mathbf{V} \mathbf{p} + \mathbf{D}(\mathbf{h})^{-1} \bm{\eta} = \mathbf{V} \mathbf{p} + \bm{\zeta},
\end{align}
where $\mathbf{p}$ and $\bm{\eta}$ are the transmit power and received noise power vectors ($\bm{\zeta} = \mathbf{D}(\mathbf{h})^{-1} \bm{\eta}$ represents the normalized noise term). Denoting the SINR vector at the receivers as $\bm{\gamma}$, we have
\begin{align}
\label{eq:defn_gamma}
\bm{\gamma} = \frac{\mathbf{p}}{\mathbf{q}}.
\end{align}
Next, we will formulate the power control problem and obtain the corresponding optimality conditions.
\subsection{Problem Formulation}
Our goal is to find the optimal power control scheme maximizing a objective function of the SINR. We consider an objective function $U(\bm{\gamma}): \mathbb{R} ^ N \rightarrow \mathbb{R}$  to be a smooth, increasing,\footnote{We assume the utility of each link should increase with SINR.} and concave\footnote{A concave utility function leads to fair assignment of network resources~\cite{Kelly:1998}.} function of the $\log$-SINR vector\footnote{We refer to such a function as $\log$-concave.} , $\log \bm{\gamma}$ (that is, $U \in C^2$, $\triangledown_{\log \bm{\gamma}} U \ge 0$, and $\bigtriangledown_{\log \bm{\gamma}}^2 U \preceq 0$). Based on this objective function, the resource allocation problem can be formulated as
\begin{align}
\label{eq:optimization_problem_not_convex}
\max_{\bm{\gamma}, \mathbf{p}, \mathbf{q}} \; &U(\bm{\gamma}) \\
\label{eq:gamma_const}
\text{subject to: } &\bm{\gamma} = \frac{\mathbf{p}}{\mathbf{q}}  \\
\label{eq:I_const}
                    &\mathbf{q} = \mathbf{V} \mathbf{p} + \bm{\zeta}.
\end{align}
Note that if $U(\bm{\gamma}) = \sum_{i = 1}^N U_i(\gamma_i)$ for concave increasing functions $U_i(\cdot)$, the optimization problem in~\eqref{eq:optimization_problem_not_convex}-\eqref{eq:I_const} turns into a network utility maximization (NUM) problem~\cite{Kelly:1998}. The optimization problem in~\eqref{eq:optimization_problem_not_convex}-\eqref{eq:I_const} is not convex due to the non-convexity of the constraint~\eqref{eq:gamma_const}. However, we show that the problem can become convex in the logarithm of the optimization variables~\cite{MChiang:2008} and~\cite{Boche:2004}. To this end, define
\begin{align}
\label{eq:defn_x_y_z}
    \mathbf{x} := \log \bm{\gamma}, \;   \mathbf{y} := \log \mathbf{p}, \;    \mathbf{z} := \log \mathbf{q},
\end{align}
leading to the following equivalent optimization problem
\begin{align}
\label{eq:optimization_problem_convex}
    \max_{\mathbf{x},\mathbf{y}, \mathbf{z}} \; &U(e^{\mathbf{x}})\\
\label{eq:z_const}
       \text{subject to :} \;        &\mathbf{z} \ge \log \left(\mathbf{V} e^{\mathbf{y}} + \bm{\zeta}\right), \\
\label{eq:x_const}
    &\mathbf{x} \le \mathbf{y - z}.
\end{align}
It can be easily verified that the optimization problems in~\eqref{eq:optimization_problem_not_convex}-\eqref{eq:I_const} and~\eqref{eq:optimization_problem_convex}-\eqref{eq:x_const} are equivalent(note that if $\mathbf{y}$ is a solution to~\eqref{eq:optimization_problem_convex}-\eqref{eq:x_const}, then the constraints~\eqref{eq:z_const} and~\eqref{eq:x_const} will be met with equality and therefore,~\eqref{eq:gamma_const} and~\eqref{eq:I_const} are satisfied). In order to prove the convexity of the optimization problem in~\eqref{eq:optimization_problem_convex}, we require the following lemma:
\begin{lemma}
\label{lemma:convexity_of_interference_constraint}
The function $f(\mathbf{x}) = \ln \left( \mathbf{a}^T e^{\mathbf{x}} + a\right)$ where $\mathbf{a} \in \mathbb{R}^{+N}$ and $a \in \mathbb{R}^+$ is convex in its domain ($\mathbf{x} \in \mathbb{R}^{N}$).
\end{lemma}
\begin{IEEEproof}
See Appendix~\ref{app:convexity_of_interference_constraint}.
\end{IEEEproof}
We prove the convexity of the optimization problem~\eqref{eq:optimization_problem_convex} in the following theorem.
\begin{theorem}
\label{theorem:convexity}
The optimization problem defined by objective function in~\eqref{eq:optimization_problem_convex}  is convex optimization if and only if $U(\cdot)$ satisfies the following
\begin{align}
\label{eq:log_concavity}
\mathbf{D}(\bm{\gamma}) \bigtriangledown_{\bm{\gamma}}^2 U \mathbf{D}(\bm{\gamma}) + \mathbf{D}(\bm{\gamma}) \mathbf{D}(\bigtriangledown_{\bm{\gamma}} U) \preceq 0.
\end{align}
\end{theorem}
\begin{IEEEproof}
Note that from Lemma~\ref{lemma:convexity_of_interference_constraint}, the feasible set defined by~\eqref{eq:z_const} and~\eqref{eq:x_const} is convex. Now the problem is convex optimization if the utility function is concave in $\mathbf{x}$, i.e., has a negative semi-definite Hessian. The condition for that is simply
\begin{align}
\label{eq:hessian_U_step_1}
\bigtriangledown_{\mathbf{x}}^2 U(e^{\mathbf{x}}) &= \left( \bigtriangledown_{\mathbf{x}} U(e^{\mathbf{x}})  \right)' =\left(\left(e^{\mathbf{x}} \right)'^T \bigtriangledown_{\bm{\gamma}} U \right)' \\
\label{eq:hessian_U_step_2}
                                & = \mathbf{D}(e^{\mathbf{x}}) \bigtriangledown_{\bm{\gamma}}^2 U \mathbf{D}(e^{\mathbf{x}}) + \mathbf{D}(e^{\mathbf{x}}) \mathbf{D}(\bigtriangledown_{\bm{\gamma}} U) \le 0.
\end{align}
where in~\eqref{eq:hessian_U_step_1} the derivative (for the Jacobian) is with respect to $\mathbf{x}$. By replacing $e^{\mathbf{x}}$ with $\bm{\gamma}$ in~\eqref{eq:hessian_U_step_2}, the requirement of the condition in~\eqref{eq:log_concavity} is proven.
\end{IEEEproof}

Note that the feasible set defined by constraints~\eqref{eq:z_const}-\eqref{eq:x_const} has none-empty interior and satisfies the qualification conditions for Slater's constraint~\cite{BoydBook} and therefore, the duality gap between the optimization problem in~\eqref{eq:optimization_problem_convex}-\eqref{eq:x_const} and its corresponding dual problem is zero. Next we use the set of Kahrun-Kush-Tucker (KKT) optimality conditions to determine the optimal solution to the problem in~\eqref{eq:optimization_problem_convex}-\eqref{eq:x_const}. The results are presented in the following theorem.

\begin{theorem}
\label{theorem:optimal_power_vector}
The optimal transmitter power vector $\mathbf{p}^*$ maximizing the objective function in~\eqref{eq:optimization_problem_convex} satisfies
the following system of nonlinear equations \begin{align}
\label{eq:optimal_power_vector}
\mathbf{p}^* = \mathbf{p}^* \frac{\bigtriangledown_{\bm{\gamma}}U \mathbf{D}(\mathbf{q}^{* })^{-1}}{\mathbf{V}^T \mathbf{D}(\bm{\gamma}^*) \bigtriangledown_{\bm{\gamma}} U \mathbf{D}(\mathbf{q}^{*})^{-1}}
\end{align}
where $\bm{\gamma}^*$ and $\mathbf{q}^*$ are the corresponding SINR and interference power vectors for the given power vector $\mathbf{p}^*$.
\end{theorem}

\begin{IEEEproof}
From KKT conditions, the gradient of the Lagrangian is zero at the optimal power vector. Following from~\eqref{eq:optimization_problem_convex}-\eqref{eq:x_const} we have
\begin{align}
\label{eq:Lagrangian_no_transmit_power_const}
L(\mathbf{x}, \mathbf{y}, \mathbf{z}, \bm{\lambda}, \bm{\mu}) = &U(e^{\mathbf{x}}) -  \bm{\lambda}^T (\mathbf{x} - \mathbf{y} + \mathbf{z}) \nonumber \\
                                                                &- \bm{\mu}^T \left(\ln \left(\mathbf{V} e^{\mathbf{y}} + \bm{\zeta} \right) - \mathbf{z} \right),
\end{align}
where $\bm{\lambda},\bm{\mu}$ represent the Lagrange multipliers. From the gradient of $L(\cdot)$ with respect to $\mathbf{x}$, $\mathbf{y}$, and $\mathbf{z}$ we obtain
\begin{align}
\label{eq:lagrangian_wrt_x}
\bigtriangledown_\mathbf{x} L = 0 &\Rightarrow
            \Rightarrow \mathbf{D}(e^{\mathbf{x}})\triangledown_{\bm{\gamma}}U = \bm{\lambda} \\
\label{eq:lagrangian_wrt_y}
\bigtriangledown_{\mathbf{y}} L = 0
&\Rightarrow \bm{\lambda} ^ T = \bm{\mu}^T \mathbf{D} \left(e^{\mathbf{z}}\right)^{-1} \mathbf{V} \mathbf{D}\left( e ^ \mathbf{y} \right) \\
\label{eq:lagrangian_wrt_z}
\bigtriangledown_{\mathbf{z}} L = 0 &\Rightarrow \bm{\lambda}^T = \bm{\mu}^T.
\end{align}
From~\eqref{eq:lagrangian_wrt_x} and~\eqref{eq:lagrangian_wrt_z}  it follows for ~\eqref{eq:lagrangian_wrt_y} that
\begin{align}
\label{eq:optimal_log_power_vector}
e^{\mathbf{y}} &= \frac{\mathbf{D}\left( e^{\mathbf{x}} \right) \bigtriangledown_{\bm{\gamma}} U}{\mathbf{V}^T \mathbf{D} \left( e^{\mathbf{z}} \right)^{-1} \mathbf{D} \left( e^{\mathbf{x}} \right) \bigtriangledown_{\bm{\gamma}}U}.
\end{align}
where by replacing $\mathbf{x}$, $\mathbf{y}$, and $\mathbf{z}$ in~\eqref{eq:optimal_log_power_vector} respectively with $\log \bm{\gamma}$, $\log \mathbf{p}$, and $\log \mathbf{q}$, the result in~\eqref{eq:optimal_power_vector} is obtained. We emphasize that the division in~\eqref{eq:optimal_log_power_vector} is element-wise.
\end{IEEEproof}

 The optimal power vector is the solution to the system of $N$ non-linear equations and $N$ variables presented in Theorem~\ref{theorem:optimal_power_vector}. As explained in the introduction, we avoid using the Newton's method to solve~\eqref{eq:optimal_power_vector} due to its computation cost for $N \gg 1$.  To this end, in Section~\ref{sec:power_control_algorithm} we formulate an algorithm based on fixed-point iteration and prove its convergence for a wide class of concave utility functions $U(\cdot)$.

 \section{Power Control Algorithm}
 \label{sec:power_control_algorithm}
 Building upon the results of Theorem~\ref{theorem:optimal_power_vector} in Section~\ref{sec:system_model}, here we propose an efficient iterative algorithm to solve~\eqref{eq:optimization_problem_not_convex}.
 We first consider the problem for the interference limited scenario, that is, we assume that compared to the interference level at the receivers,  $\bm{\zeta} \simeq \mathbf{0}$ and can be neglected from our formulation\footnote{Note that this assumption is equivalent to having no maximum power constraints on the wireless nodes. With no power constraint on the transmitters, the wireless nodes can transmit at very high power leading to interference levels far greater than the thermal noise power, i.e., relative to interference level we can assume $\bm{\zeta} \simeq \mathbf{0}$.}. Later on we will extend the results from the interference limited scenario to the scenario with  $\bm{\zeta} > \mathbf{0}$ (which includes maximum and minimum transmit power constraints).

 \subsection{Power Control under Interference Limited Circumstances}
 \label{sec:power_control_algorithm_IoT}
Assuming $\bm{\zeta} = \mathbf{0}$ and defining $\mathbf{S} :=  \mathbf{D}(\bm{\gamma}) \mathbf{V}$, from~\eqref{eq:defn_q} and~\eqref{eq:defn_gamma} it follows that
 \begin{align}
  \label{eq:p_eigenvector_of_S}
 \mathbf{S} \mathbf{p} &= \mathbf{p}.
 \end{align}
 Note that $\mathbf{p} \in \mathbb{R}^{N+}$ is the (right) eigenvector of the corresponding matrix $\mathbf{S}$ with an eigenvalue of $1$. In the following lemma, we discuss the left and right eigenvectors of  $\mathbf{S}$ and prove that $\rho[\mathbf{S}] = 1$ where $\rho[\cdot]$ represents the spectral radius of the matrix.
\begin{lemma}
\label{lemma:p_alpha_ev_S}
For any $\mathbf{p} \ge 0$ and $\mathbf{S}$ defined in~\eqref{eq:p_eigenvector_of_S}, $\mathbf{p}$ is the unique eigenvector corresponding to the spectral radius of $\mathbf{S}$ where $\rho(\mathbf{S}) = 1$. Furthermore, the left eigenvector of $\mathbf{S}$, namely, $\bm{\alpha}$,  corresponding to the spectral radius of $\mathbf{S}$ satisfies the following at the optimal power vector.
\begin{align}
\label{eq:defn_alpha}
    \bm{\alpha} = \mathbf{D}\left( \mathbf{q} \right) ^ {-1} \bigtriangledown_{\bm{\gamma}} U.
\end{align}
\begin{IEEEproof}
See Appendix~\ref{app:p_alpha_ev_S}.
\end{IEEEproof}
\end{lemma}
%
%
From Lemma~\ref{lemma:p_alpha_ev_S} and~\eqref{eq:optimal_power_vector} we have the following equality at the optimal power vector
\begin{align}
\label{eq:defn_phi}
&\mathbf{p}^* = \mathbf{p}^* \frac{\bm{\alpha}^*}{\mathbf{S}^T \bm{\alpha}^*} =  \mathbf{p}^* \bm{\phi}(\mathbf{p}^*), \; \text{with } \; \bm{\phi}(\mathbf{p}) = \frac{\bm{\alpha}}{\mathbf{S}^T \bm{\alpha}},
\end{align}
and $\bm{\alpha}$ defined according to~\eqref{eq:defn_alpha} (note that for an arbitrary $\mathbf{p}$, $\bm{\alpha}$ is not necessarily the left eigenvector of $\mathbf{S}$). Equation~\eqref{eq:defn_phi} suggests that the optimal power vector$\mathbf{p}^*$ is a fixed point of the vector function $\mathbf{p} \bm{\phi}(\mathbf{p})$ (note that $\bm{\phi}(\mathbf{p}^*) = 1$). This motivates us to propose a fixed-point iteration based on~\eqref{eq:defn_phi} to solve the system of non-linear equations in Theorem~\ref{theorem:optimal_power_vector}.

 From Banach's theory on fixed-point iterations~\cite{Hizler:2001}, the local convergence for smooth fixed-point functions relates to the spectral radius of the Jacobian matrix at the fixed-point. Specifically, if the Jacobian matrix has a spectral radius smaller than $1$, the fixed-point iteration converges\footnote{In fact, Banach's theory expresses the convergence conditions in terms of the~\emph{norm} of the Jacobian. However, for a matrix $\mathbf{M}$ and any $\epsilon > 0$, there is a definition of norm $\parallel \cdot \parallel$ such that $\rho(\mathbf{M}) \le \parallel \mathbf{M} \parallel \le \rho(\mathbf{M}) + \epsilon$ (See~\cite{OtegaBook}). Therefore, the condition can be extended to the spectral radius instead.}.

Before presenting the fixed-point iteration and proving the convergence of the algorithm, we require the Jacobian matrix of the vector function $\bm{\phi}(\cdot)$. The following lemmas determine the Jacobian of $\bm{\phi}(\cdot)$ at the optimal power vector and some of its properties. To keep the formulation simple, for the rest of the paper, we define and use the vector $\mathbf{r} := \frac{\mathbf{p}}{\bm{\alpha}}$ as the ratio the power vector over $\bm{\alpha}$ defined in~\eqref{eq:defn_alpha}.
\begin{lemma}
\label{lemma:jacobian_phi}
The Jacobian of vector function $\bm{\phi}(\cdot)$ at $\mathbf{p} = \mathbf{p}^*$ satisfies
\begin{align}
 \mathbf{D}(\bm{\alpha})  \bm{\phi}'(\mathbf{p}) &=\left[\tilde{\mathbf{S}}^T \left(\mathbf{D}(\mathbf{q})^{-1} \bigtriangledown_{\bm{\gamma}}^2 U \mathbf{D}(\mathbf{q})^{-1} + 2\mathbf{D} \left(\mathbf{r}\right)^{-1} \right) \tilde{\mathbf{S}} \right. \nonumber \\
&\left.  - \left(\tilde{\mathbf{S}}^T\mathbf{D} \left(\mathbf{r}\right)^{-1} + \mathbf{D} \left(\mathbf{r}\right)^{-1} \tilde{\mathbf{S}}  \right) \right],  \\
\label{eq:r_and_S_tilde}
\text{with: }\; &\mathbf{r} = \frac{\mathbf{p}}{\bm{\alpha}}, \; \tilde{\mathbf{S}} := \mathbf{I} - \mathbf{S},
\end{align}
 and $\mathbf{I}$ is as the $N \times N$ identity matrix.
\end{lemma}
\begin{IEEEproof}
Ref. to Appendix~\ref{app:jacobian_phi}.
\end{IEEEproof}
\begin{lemma}
\label{lemma:sign_of_second_term}
The matrix
\begin{align}
\mathbf{D}\left(\mathbf{r}\right) ^ {\frac{1}{2}} \left(\tilde{\mathbf{S}}^T \mathbf{D} \left( \mathbf{r} \right) ^{-1} \tilde{\mathbf{S}}  - \tilde{\mathbf{S}}^T \mathbf{D} \left( \mathbf{r} \right)^{-1} - \mathbf{D} \left( \mathbf{r} \right)^{-1} \tilde{\mathbf{S}} \right) \mathbf{D}\left(\mathbf{r}\right) ^ {\frac{1}{2}}, \nonumber
\end{align}
is negative semi-definite, has a spectral radius of $2$, and has exactly one eigenvalue of $0$ corresponding to the eigenvector of $\left( \bm{\alpha} \mathbf{p} \right)^{\frac{1}{2}}$.
\end{lemma}
\begin{IEEEproof}
See Appendix~\ref{app:sign_of_second_term}.
\end{IEEEproof}

\begin{lemma}
\label{lemma:sign_of_jacobian}
Assume the $\log$-concave utility function $U(\cdot)$ satisfies the following\footnote{Equivalently, $\log \bigtriangledown_{\bm{\gamma}} U$ is Lipschitz continuous in $\log \bm{\gamma}$ with a Lipschitz constant $B$.}
\begin{align}
\label{eq:log_concavity_bound}
\rho\left[ \mathbf{D}\left(\frac{\bm{\gamma}}{\bigtriangledown_{\bm{\gamma}} U }\right) ^ {\frac{1}{2}} \bigtriangledown_{\bm{\gamma}} ^ 2 U \mathbf{D}\left(\frac{\bm{\gamma}}{\bigtriangledown_{\bm{\gamma}} U }\right) ^ {\frac{1}{2}} \right] \le B, \; \forall \bm{\gamma} \ge 0.
\end{align}
Then, at the optimal power vector $\mathbf{p} = \mathbf{p}^*$, we have
\begin{align}
\mathbf{D}\left(\mathbf{r} \right) ^ {\frac{1}{2}} \mathbf{D}\left(\bm{\alpha}\right) &\bm{\phi}'(\mathbf{p}) \mathbf{D} \left(\mathbf{r} \right)^{\frac{1}{2}} \preceq 0, \nonumber \\
 \rho\left[ \mathbf{D}\left(\mathbf{r} \right) ^ {\frac{1}{2}} \mathbf{D}\left(\bm{\alpha}\right) \bm{\phi}'(\mathbf{p}) \mathbf{D} \left(\mathbf{r} \right)^{\frac{1}{2}} \right] &= \rho \left[ \mathbf{D}(\mathbf{p}) \bm{\phi}'(\mathbf{p}) \right]  = \le 4B - 2 . \nonumber
\end{align}

\end{lemma}
\begin{IEEEproof}
See Appendix~\ref{app:sign_of_jacobian}.
\end{IEEEproof}
Now we can propose the following fixed point iterations which converges to the optimal solution of~\eqref{eq:optimization_problem_not_convex} for a class of utility functions described in Lemma~\ref{lemma:sign_of_jacobian}.

\begin{theorem}
\label{theorem:fixed_point_iterations_IoT}
Consider the optimization problem~\eqref{eq:optimization_problem_not_convex} with a smooth $\log$-concave utility function $U(\bm{\gamma})$ satisfying~\eqref{eq:log_concavity_bound}. Then, $\forall \; \theta \in (0, \theta_{max})$, with $\theta_{max} = \frac{1}{2B - 1}$, the following iterations converges to (an) optimal solution of~\eqref{eq:optimization_problem_not_convex}
\begin{align}
\label{eq:fixed_point_iterations}
    \mathbf{p}_{k} = \bm{\phi}_{\theta}(\mathbf{p}_{k - 1}) = \theta \mathbf{p}_{k - 1}\bm{\phi}(\mathbf{p}_{k - 1}) + (1 - \theta) \mathbf{p}_{k - 1}, \\
    \text{with arbitrary } \mathbf{p}_0 \in \mathbb{R}^{N+}. \nonumber
\end{align}
\end{theorem}
\begin{IEEEproof}
The proof is complete if we show that at the optimal point $\mathbf{p}$, $\rho \left[ \bm{\phi}_{\theta}'(\mathbf{p}) \right] < 1$~\cite{OtegaBook,Hizler:2001}. It simply follows that
\begin{align}
\label{eq:phi_theta_jacobian}
\rho \left[\left(\bm{\phi}_{\theta}(\mathbf{p})\right)' \right] &= \rho \left[ \theta \mathbf{D}(\mathbf{p}) \bm{\phi}'(\mathbf{p}) + \mathbf{I} \right] \nonumber \\
&=  \rho \left[ \theta \mathbf{D}\left(\mathbf{r} \right) ^ {\frac{1}{2}} \mathbf{D}\left(\bm{\alpha}\right) \bm{\phi}'(\mathbf{p}) \mathbf{D} \left(\mathbf{r} \right)^{\frac{1}{2}} + \mathbf{I} \right],
\end{align}
where~\eqref{eq:phi_theta_jacobian} readily follows from the definition of $\mathbf{r}$ in Lemma~\ref{lemma:jacobian_phi}\footnote{Note that for any two matrices $\mathbf{A}$ and $\mathbf{B}$ where $\mathbf{B}$ is invertible, $\rho [\mathbf{A}] = \rho[\mathbf{B}^{-1} \mathbf{A} \mathbf{B}]$ and $\rho [\mathbf{AB}] = \rho [\mathbf{BA}]$.}. From Lemma~\ref{lemma:sign_of_jacobian} we know that the matrix in~\eqref{eq:phi_theta_jacobian} has a largest positive eigenvalue of $1$ and a largest (in magnitude) negative eigenvalue of at most $4(B - 1)$. Therefore any $\theta < \frac{1}{2B - 1}$ guarantees that $\rho\left[ \bm{\phi}_{\theta}(\mathbf{p})' \right] \le 1$. On the other hand, from Lemmas~\ref{lemma:sign_of_second_term} and~\ref{lemma:sign_of_jacobian} we know that for matrix $\bm{\phi}'(\mathbf{p})$, $\mathbf{p}$ is the~\emph{only} eigenvector for the~\emph{only} eigenvalue of $1$ , i.e., $\rho \left[\bm{\phi}_{\theta}'(\mathbf{p}) \right] = 1$. However, this does not contradict with the convergence conditions since $\mathbf{p}$ is the optimal power vector. The reason is that if at iteration $k$, $\mathbf{p}_k$ is close to the optimal power vector, then the projection of the error onto the direction of the optimal power vector remains intact while the other components of the error are attenuated based on the other eigenvalues of $\bm{\phi}(\mathbf{p})'$\footnote{Also, note that in the interference above thermal scenario, if $\mathbf{p}$ is an optimal solution, then $a \mathbf{p}$ for $a > 0$ is also optimal, i.e., the optimal solution is a line in $\mathbb{R}^{N+}$ and is~\emph{not} unique.}.  Hence, the algorithm converges to the optimal power vector $\mathbf{p}$ (or a positive multiple of $\mathbf{p}$).
\end{IEEEproof}

From theorem~\eqref{theorem:fixed_point_iterations_IoT} we realize that while the iterations converge to the optimal power vector, the limit is not unique. In the following corollary we propose a modified fixed-point iteration which converges to a unique optimal power vector with finite norm.
\begin{corollary}
\label{corollary:finite_norm_fixed_point}
The following fixed point iteration converges to a solution $\mathbf{p}$ for the problem in~\eqref{eq:optimization_problem_not_convex}-\eqref{eq:I_const} where $\parallel \mathbf{p} \parallel_2 = 1$.
\begin{align}
\label{eq:finite_norm_fixed_point}
\mathbf{p}_{k} = \bm{\phi}_{\theta} \left( \frac{\mathbf{p}_{k - 1}}{\parallel \mathbf{p}_{k - 1} \parallel_2} \right).
\end{align}
\end{corollary}
\begin{IEEEproof}
See Appendix~\ref{app:finite_norm_fixed_point}.
\end{IEEEproof}


The algorithm proposed in Theorem~\ref{theorem:fixed_point_iterations_IoT} requires iterative evaluations of function $\bm{\phi}(\cdot)$, which requires $\mathcal{O}(N^2)$ calculations. Therefore, the fixed-point iterations based on~\eqref{eq:defn_phi} will have a complexity of $\mathcal{O}(N^2)$. We would like to emphasize that this level of complexity is less than both gradient-based and Newton's iterations proposed in the literature and as a result, leads to faster convergence for $N \gg 1$.
In Section~\ref{sec:power_control_non_IoT}, we extend the results of Theorem~\ref{theorem:fixed_point_iterations_IoT} to the scenario with non-zero additive noise and show that the algorithm converges for an even larger set of utility functions.

\subsection{Power Control with non-zero Noise Power}
\label{sec:power_control_non_IoT}
In this scenario, we assume $\bm{\zeta} > 0$, and consider minimum and maximum power constraints at the wireless transmitters, i.e., for wireless user $i$, $p_{min, i} \le p_i \le p_{max, i}$ (Remember assuming non-zero noise power~\emph{but} no power constraint at the transmitters leads to the interference limited scenario). We start by deriving the optimality conditions for a power constrained problem. We prove the results for the case where each wireless transmitter has a maximum transmit power constraint. The analysis for the scenario with minimum power constraints is very similar.

In order to proceed, we need to add a new constraint to the optimization problem in~\eqref{eq:optimization_problem_convex}-\eqref{eq:x_const}, i.e., $\mathbf{y} = \ln \mathbf{p} \le \ln \mathbf{p}_{max}$. In the following theorem, we provide the necessary and sufficient conditions for the optimality of a solution to the power-constrained problem.
\begin{theorem}
\label{theorem:optimality_condition_constrained}
The optimal power vector $\mathbf{p}^*$, maximizing the optimization problem~\eqref{eq:optimization_problem_not_convex} under the transmit power constraint $\mathbf{p} \le \mathbf{p}_{max}$, satisfies the following equation
\begin{align}
\label{eq:optimal_power_vector transmit_power_constraint}
\mathbf{p}^* = \min \left( \mathbf{p}_{max}, \mathbf{p}^* \bm{\phi}(\mathbf{p}^*) \right).
\end{align}
\end{theorem}
\begin{IEEEproof}
 Without loss of generality assume $p_{max, i} = 1$ for all nodes (for $p_{max, i} \neq 1$,  normalize row $i$ of $\mathbf{V}$ and also $\zeta_i$ to $p_{max, i}$), leading to the constraint $\mathbf{y} \le 0$. Since the new constraint is convex, we can use the KKT conditions to identify the optimal solution. Define the new Lagrangian to be $L_c(\cdot)$ and consider $\bm{\beta} \ge 0$ as the the new dual variable. The gradient of Lagrangian with respect to $\mathbf{x}$ and $\mathbf{z}$ is similar to~\eqref{eq:lagrangian_wrt_x} and~\eqref{eq:lagrangian_wrt_z} respectively. From the gradient of $L_c(\cdot)$ with respect to $\mathbf{y}$  it follows that
\begin{align}
\bigtriangledown_{\mathbf{y}} L_c  = 0 \Rightarrow \bm{\lambda}^T - \bm{\beta}^T = \bm{\lambda}^T\mathbf{D}\left(e^{\mathbf{z}}\right) ^ {-1} \mathbf{V} \mathbf{D}\left(e^{\mathbf{x}}\right)
\end{align}
where after replacing $\mathbf{x}$, $\mathbf{y}$, and $\mathbf{z}$ respectively with $\bm{\gamma}$, $\mathbf{p}$, and $\mathbf{q}$ we obtain
\begin{align}
\mathbf{p}^* = \frac{\bm{\alpha}^*-\frac{\bm{\beta}^*}{\mathbf{p}^*}}{\mathbf{S}^{*T} \bm{\alpha}^*}.
\end{align}
From above, if $\beta_i = 0$ then $p_i^* = \left( \bm{\phi}(\mathbf{p}^*)_i \right) p_i^*$. If $\beta_i > 0$, by the optimality conditions we must have $y_i^* = 0$ which leads to $p_i^* = 1$. On the other hand, $\alpha_i^* - \beta_i^* / p_i^* < \alpha_i^*$ and therefore, $p_i^* (\bm{\phi}(\mathbf{p}^*))_i > 1$ which satisfies $p_i^* = \min \left( 1, p_i^* \left( \bm{\phi} ( \mathbf{p}^* ) \right)_i \right)$.
\end{IEEEproof}
For the problem with minimum constraints on the transmit power of wireless nodes a similar result can be obtained.
\begin{theorem}
\label{theorem:min_max_power_constraint_optimality}
The solution to the optimization power control problem presented in~\eqref{eq:optimization_problem_not_convex} with the maximum and minimum power constraint ($\mathbf{p}_{min} \le \mathbf{p} \le \mathbf{p}_{max}$) satisfies the following
\begin{align}
    \mathbf{p}^* = \max \left( \mathbf{p}_{min}, \min \left( \mathbf{p}_{max}, \mathbf{p}^* \bm{\phi}(\mathbf{p}^*) \right) \right).
\end{align}
\end{theorem}
\begin{IEEEproof}
The sketch of the proof is similar to that of Theorem~\ref{theorem:optimality_condition_constrained} and can be easily repeated by the reader.
\end{IEEEproof}

The power control algorithm converging to the optimal solution of~\eqref{eq:optimization_problem_not_convex} under maximum power constraints is similar to that of interference limited networks. Before proceeding with the proof of convergence, we present the Jacobian of $\bm{\phi}_{\theta}(\cdot)$ (defined in~\eqref{eq:fixed_point_iterations}) at the optimal power vector in the following lemma.
\begin{lemma}
\label{lemma:jacobian_phi_non_IoT}
For the optimization problem in~\eqref{eq:optimization_problem_not_convex} with $\bm{\zeta} > 0$ and maximum power constraints, the Jacobian matrix of $\bm{\phi}_{\theta}(\cdot)$ at the optimal solution satisfies
\begin{align}
\bm{\phi}_{\theta}'(\mathbf{p}) = \mathbf{D}(\bm{\phi}(\mathbf{p})) \bm{\phi}_{\theta, \bm{\zeta} = \mathbf{0}}'(\mathbf{p}) + \mathbf{J},
\end{align}
where $\bm{\phi}_{\theta, \bm{\zeta} = \mathbf{0}}'(\mathbf{p})$ is the Jacobian at the optimal solution \emph{as if} $\bm{\zeta} = 0$ and $\mathbf{J}$ is a matrix with the property that for any $\mathbf{v} \in \mathbb{R}^N$ and any $i$ that $\left(\bm{\phi}(\mathbf{p})\right)_i = 1$, $\left( \mathbf{J} \mathbf{v} \right)_i = 0$.
\end{lemma}
\begin{IEEEproof}
See Appendix~\ref{app:jacobian_phi_non_IoT}.
\end{IEEEproof}

In the next theorem, we present the fixed-point iteration that converges to the optimal power vector under maximum transmit power constraints at the wireless transmitters.

\begin{theorem}
\label{theorem:fixed_point_iterations_non_IoT}
The following fixed-point iteration converges to the global optimum of power control problem defined in Theorem~\ref{theorem:fixed_point_iterations_IoT} under maximum power constraints at each node,
\begin{align}
\label{eq:fixed_point_under_max_power}
\mathbf{p}_{k} =  \min \left(\mathbf{p}_{max}, \bm{\phi}_{\theta}(\mathbf{p}_{k - 1})  \right) = \min \left(\mathbf{1}, \bm{\phi}_{\theta}(\mathbf{p}_{k - 1}) \right).
\end{align}
\end{theorem}
\begin{IEEEproof}
First note that from Theorem~\ref{theorem:optimality_condition_constrained}, any fixed point of~\eqref{eq:fixed_point_under_max_power} is an optimal solution to the optimization problem defined in~\eqref{eq:optimization_problem_not_convex} under maximum transmit power constraints. Since the function in~\eqref{eq:fixed_point_under_max_power} is not differentiable at the optimal power vector, the proof is slightly different from that of Theorem~\ref{theorem:fixed_point_iterations_IoT}. Without loss of generality assume $(\bm{\phi}_{\theta}(\mathbf{p}))_i > 1$ (or equivalently, $(\bm{\phi}(\mathbf{p}))_i > 1$ ) for $ i \in \{1, 2, ..., n\}$ and $(\bm{\phi}_{\theta}(\mathbf{p}))_i  = 1$ for $i \in \{n + 1, n + 2, ..., N\}$. We know that $\bm{\phi}(\mathbf{p})$ is continuous in its domain and therefore, for any $\epsilon > 0$, we can find $r > 0$ small enough such that $\sup_{\mathbf{p}' \in B(\mathbf{p}, r)} \parallel \bm{\phi}_{\theta}(\mathbf{p}') - \bm{\phi}_{\theta}(\mathbf{p}) \parallel < \epsilon $ where $B(\mathbf{p}, r)$ is an open ball of radius $r$ (in $\mathbb{R}^N$) centered at $\mathbf{p}$. Choose the $r$ corresponding to an $\epsilon < \min_{1 \le i \le n} (\bm{\phi}_{\theta}(\mathbf{p}))_i$.

Now suppose at iteration $k$ we have $\mathbf{p}_k = \mathbf{p} + \bm{\epsilon}_k$ with $\bm{\epsilon}_k \in B(\mathbf{0}, r)$. Define $\mathbf{A}_n$ as a diagonal matrix with $(\mathbf{A}_n)_{i, i} = 0$ for $i \le n$ and $(\mathbf{A}_n)_{i, i} = 1$ for $i > n$. It follows for the power vector at iteration $k + 1$ that
\begin{align}
\label{eq:error_at_k_plus_1_s1}
    \mathbf{p}_{k + 1}  &= \min\left( \mathbf{1}, \bm{\phi}_{\theta}(\mathbf{p} + \bm{\epsilon}_k) \right) \simeq \left(\mathbf{I} - \mathbf{A}_n \right) \mathbf{1} \nonumber \\
    &+ \mathbf{A}_n \left( \mathbf{p} + \bm{\phi}_{\theta}'(\mathbf{p}) \bm{\epsilon}_k \right) = \mathbf{p} + \mathbf{A}_n \bm{\phi}_{\theta}'(\mathbf{p}) \bm{\epsilon}_k.
\end{align}
 Note that~\eqref{eq:error_at_k_plus_1_s1} comes from the continuity and linear approximation of $\bm{\phi}_{\theta}(\cdot)$ in the vicinity of $\mathbf{p}$. From~\eqref{eq:error_at_k_plus_1_s1} and Lemma~\ref{lemma:jacobian_phi_non_IoT}, we have
\begin{align}
\label{eq:error_at_k_plus_1_s3}
\bm{\epsilon}_{k + 1} &= \mathbf{p}_{k + 1} - \mathbf{p} \simeq \mathbf{A}_n \bm{\phi}_{\theta}'(\mathbf{p}) \bm{\epsilon}_k  = \mathbf{A}_n \bm{\phi}_{\theta, \bm{\zeta} = \mathbf{0}}'(\mathbf{p}) \bm{\epsilon}_{k},
\end{align}
where~\eqref{eq:error_at_k_plus_1_s3} is true since $\mathbf{A}_n \mathbf{J} = \mathbf{0}_{N \times N}$ and $\mathbf{A}_n \mathbf{D}\left(\bm{\phi}(\mathbf{p}) \right)= \mathbf{A}_n$. Furthermore, from the proof of Lemma~\ref{lemma:jacobian_phi_non_IoT} we have $\rho\left[\mathbf{A}_n \bm{\phi}_{\theta, \bm{\zeta} = \mathbf{0}}'(\mathbf{p}) \right] \le \rho\left[ \bm{\phi}_{\theta, \bm{\zeta} = \mathbf{0}}'(\mathbf{p}) \right] < 1$ and therefore $\parallel \bm{\epsilon}_{k + 1} \parallel \le \rho \parallel \bm{\epsilon}_{k} \parallel$ for some consistent norm function $\parallel \cdot \parallel$. This concludes the proof of local convergence of the iterations.
\end{IEEEproof}
Based on a similar approach, we can propose the following fixed point iterations for the scenario with both maximum and minimum power constraints.
\begin{theorem}
\label{theorem:fixed_point_iterations_min_max_power}
For any $U(\cdot)$ and $\theta$ satisfying the conditions of Theorem~\ref{theorem:fixed_point_iterations_IoT}, the following iteration converges to the optimal solution of problem~\eqref{eq:optimization_problem_not_convex} with the minimum and maximum power constraints $\mathbf{p}_{min}$ and $\mathbf{p}_{max}$
\begin{align}
\mathbf{p}_{k} = \max\left( \mathbf{p}_{min}, \max \left( \mathbf{p}_{max}, \bm{\phi}_{\theta}(\mathbf{p}_{k - 1}) \right) \right) \nonumber.
\end{align}
\end{theorem}
\begin{IEEEproof}
The proof is similar to that of Theorem~\ref{theorem:fixed_point_iterations_non_IoT} and is not repeated here due to space constraints.
\end{IEEEproof}
The convergence of fixed-point iterations under maximum and minimum power constraints at the wireless transmitters depends on the stability of the Jacobian matrix $\bm{\phi}_{\theta, \bm{\zeta} = \mathbf{0}}'(\mathbf{p})$ which is, in turn, guaranteed if the utility function satisfies the conditions mentioned in Theorem~\ref{theorem:fixed_point_iterations_IoT} and $\theta$ is chosen small enough. The following corollary extends the proof of convergence to all $\log$-concave utility functions for an optimization of the form of~\eqref{eq:optimization_problem_not_convex}-~\eqref{eq:I_const} under minimum and maximum power constraint.
\begin{corollary}
\label{corollary:max_min_power}
For any strictly increasing $\log$-concave utility function $U(\cdot)$, the fixed point iteration defined in Theorem~\ref{theorem:fixed_point_iterations_min_max_power} converges to a point which satisfies the KKT conditions for the optimization problem~\eqref{eq:optimization_problem_not_convex} under maximum and minimum transmit power constraints.
\end{corollary}
\begin{IEEEproof}
Since the utility function is smooth, over compact set $\mathcal{S} = \left\{ \mathbf{p}: \mathbf{p}_{min} \le \mathbf{p} \le \mathbf{p}_{max} \right\}$, we have $\bm{\phi}_{\theta}'(\mathbf{p}) < B_{\mathcal{S}}$ for some finite positive $B_{\mathcal{S}}$. Then, for any $\theta < \theta_{max} < \frac{1}{2B_{\mathcal{S}} - 1}$ the power control algorithm is locally convergent.
\end{IEEEproof}
\subsection{Asynchronous Power Updates}
As a final note, a property that may be of implementation benefits is that the proposed power control algorithm does not require synchronous power updates\footnote{Asynchronous updates is defined as the process of calculating the new value of variable $(\mathbf{x}_{k + 1})_i$ based on $\left( \mathbf{x}_{k} \right)_j$ for $j \ge i$, and $\left(\mathbf{x}_{k + 1} \right)_j$ for $j > i$.}. In fact, the convergence speed might be faster under asynchronous power updates. The proof of convergence under asynchronous power updates follows a general property of fixed point iterations. Consider $\mathbf{x} = \mathbf{f}(\mathbf{x})$ a converging iteration to a point $\mathbf{x}_0$ with $\rho \left[ \mathbf{f}'(\mathbf{x}_0) \right] < 1 $.  The asynchronous power update process can be formulated as follows
\begin{align}
\label{eq:asynch_1}
\mathbf{x}_{k + 1} &= \mathbf{g}_N(\mathbf{g}_{N - 1}(...(\mathbf{g}_1(\mathbf{x}_k))...)) \nonumber \\
\text{with:} \nonumber \\
 \mathbf{g}_i(\mathbf{x}) &= \mathbf{D}(\mathbf{e}_i) \mathbf{f}(\mathbf{x}) + \left(\mathbf{I} - \mathbf{D}(\mathbf{e}_i) \right) \mathbf{x}
\end{align}
and $\mathbf{e}_i$ is a vector with a single non-zero element $\left(\mathbf{e}_i \right)_i = 1$. The new fixed point iteration defined in~\eqref{eq:asynch_1} converges if the Jacobian matrix is bounded by $1$. It follows that
\begin{align}
\left( \mathbf{g}_N(\mathbf{g}_{N - 1}(...(\mathbf{g}_1(\mathbf{x}_0))...)) \right)' = \mathbf{g}_N'(\mathbf{x}_0) \mathbf{g}_{N - 1}'(\mathbf{x}_0)...\mathbf{g}_1'(\mathbf{x}_0) \nonumber \\
 = \prod_{i = 1} ^ {N} \left(\mathbf{D}(\mathbf{e}_i) \mathbf{f}'(\mathbf{x}_0) + \mathbf{I} - \mathbf{D}(\mathbf{e}_i)\right)  = \mathbf{f}'(\mathbf{x}_0), \nonumber
\end{align}
suggesting that the asynchronous methods also converges.

\subsection{Distributed Power Control and Implementation Issues}
In order to solve the optimization problem~\eqref{eq:optimization_problem_convex}, the destination nodes need to know the interference matrix. This represents a significant signaling overhead of $\mathcal{O}(N^2)$ which might not be practical in dense wireless networks. Here we show that through distributed implementation, the signaling overhead can be reduced to $\mathcal{O}(N)$. From the fixed-point iterations in Theorems~\ref{theorem:fixed_point_iterations_IoT} and~\ref{theorem:fixed_point_iterations_non_IoT} it is observed that at a given iteration, the destination of wireless transmitter $i$ requires the values of $\alpha_i$ and $\left(\mathbf{S}^T \bm{\alpha}\right)_i$ to find the new transmit power. On the other hand from Lemma~\ref{lemma:p_alpha_ev_S}, the value of $\alpha_i$ depends on the gradient of the utility function as well as the received interference power, $q_i$. Assuming the receiver can estimate the value of $q_i$, knowledge of $(\bigtriangledown_{\bm{\gamma}} U)_i$ is required for the evaluation of $\alpha_i$. In order to evaluate $\left(\bigtriangledown_{\bm{\gamma}} U(\bm{\gamma}) \right)_i$, global knowledge of the SINR vector $\bm{\gamma}$ is necessary. Therefore, we assume that at the beginning of each iteration, every destination node broadcasts the SINR for the corresponding wireless transmitter.

Furthermore, for $\left( \mathbf{S}^T \bm{\alpha} \right) $ it follows that
\begin{align}
\left(\mathbf{S}^T \bm{\alpha}\right)_i = \sum_{j = 1}^{N} S^T_{i,j} \alpha_j =  \sum_{j = 1} ^ {N} V_{j,i} \gamma_j \alpha_j  = \sum_{j = 1} ^ {N} H_{j, i} \kappa_j, \nonumber
\end{align}
where $\kappa_i = \frac{\gamma_i \alpha_i}{h_i}$. Therefore, in order to evaluate $\left( \mathbf{S}^T \bm{\alpha} \right)_i$ at each destination node, global knowledge of vector $\bm{\kappa}$ at each iteration, and~\emph{local} knowledge of channel gains is required. Note that this local knowledge of channel gains includes channels gains from wireless transmitter to other destination nodes in the network which can be provided through channel feedback or channel reciprocity assumption in the case of TDD channel access.

Algorithm 1 provides a distributed implementation of the proposed power control and converges to the optimal solution of~\eqref{eq:optimization_problem_not_convex}.
\begin{figure}[h]
\begin{algorithmic}[1]
\State{Variables at user $i$, iteration $k$: $p_i^{(k)}$, $q_i^{(k)}$, $\gamma_i^{(k)}$, $\alpha_i^{(k)}$, $\phi_i^{(k)}$, and $\kappa_i^{(k)}$.}
\State{Initialize $p_i^{(0)} \in (0, p_{max}]$, $\theta \in (0, 1]$, $M \in \mathbb{N}$, and $k, c = 1$}
\Repeat
    \State{Estimate received interference power $q_i^{(k)}$}
    \State{Broadcast $\gamma_i^{(k)} = \gets \frac{p^{(k)}}{q_i^{(k)}}$}
    \State{Calculate $\alpha_i^{(k)} \gets \frac{\left( \bigtriangledown_{\bm{\gamma}} U (\bm{\gamma}^{(k)})\right)_i }{q_i^{(k)}}$}
    \State{Broadcast $\kappa_i^{(k)} \gets \frac{\gamma_i^{(k)} \alpha_i^{(k)}}{h_i}$}
    \State{Calculate $\phi_i^{(k)} \gets \frac{\alpha_i^{(k)}}{\sum_{j = 1} ^ {N} H_{j,i} \kappa_j^{(k)}}$}
    \State{Update power: $p_i^{(k + 1)} \gets \min\left(p_{max}, p_i^{(k)} \phi_i^{(k)}\right)$}
    \State{Calculate error : $\epsilon^{(k + 1)} \gets \mid p_i^{(k + 1)} - p_i^{(k)} \mid$}
    \If{$k > cM$}
      \State{$\theta \gets \frac{\theta}{2}$}
      \State{$c \gets c + 1$}
    \EndIf
    \State{$k \gets k + 1$}
\Until{$\epsilon^{(k + 1)} < tol.$}
\end{algorithmic}
\caption*{Algorithm 1 - Distributed Power Control: General Utility Function}
\end{figure}
Note that implementation of Algorithm 1 requires $2N$ messages broadcasted at each iteration. More importantly, evaluation of the power vector depends on the local channel gains which leads to a signaling complexity of $\mathcal{O}(N)$ in between the destination nodes.
From Algorithm 1, two phases of signaling are required at each iteration: During the first round the destination node(s) broadcast the SINR vector to allow other destinations to evaluate the gradient of the utility function, and subsequently, the value of $\bm{\alpha}$. The second round involves broadcasting the value of $\bm{\alpha}$. This information is used to determine how much to increase/decrease the transmit power for the next iteration. Note that an assumption of NUM for the optimization problem obviates the need for the first round of signaling (each node can find the derivative of its own utility individually based on its own SINR and interference power). This suggests an overall $N$ signaling messages per iteration.
Finally, in Algorithm 1, the value of $\theta$ is adaptive, that is, after each $M$ iterations, $\theta$ becomes smaller. In general, the initial value of $\theta$ and $M$ could be tweaked to obtain better convergence properties than that of fixed $\theta$.

\section{Numerical Results}
\label{sec:numerical_results}
\begin{figure}[t]
\centering
\includegraphics[width = 3.5in]{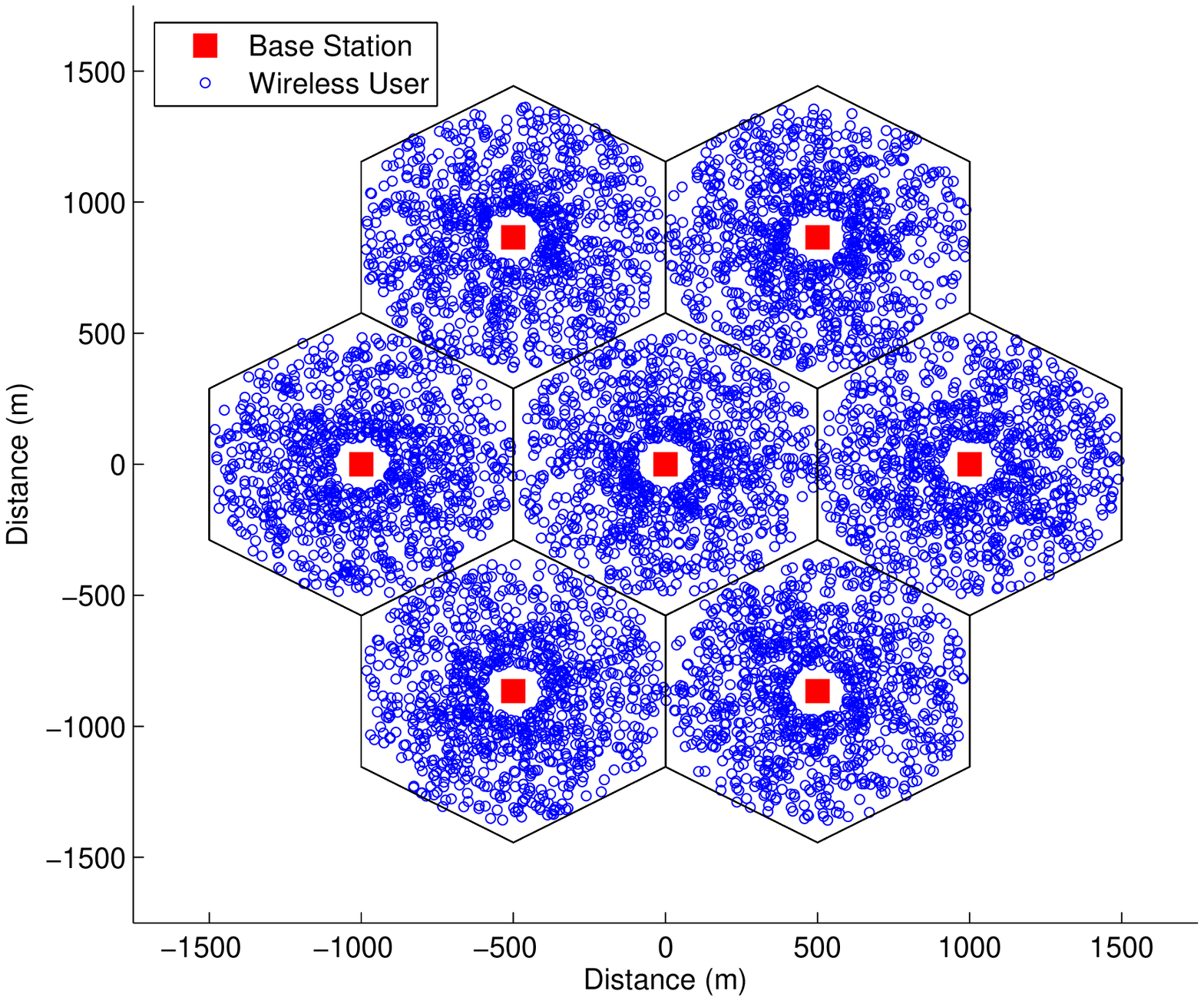}
\caption{Distribution of wireless node in a $7$-cell network topology.}
\label{fig:network_topology}
\end{figure}

\setlength{\textfloatsep}{5 pt}
In order to validate the theory developed and illustrate the performance of our proposed power control algorithm we present the results of simulations based on popular system models. We consider a seven-cell cellular topology with hexagonal cells where, within each cell, $N_{cell}$ wireless users are connected to the central base station. We consider a dense interference network where, due to the short distance between the base-stations, inter-cell interference imposes significant performance degradation. We adopted the IEEE 802.16-based channel models for fixed wireless network with intermediate terrain and path loss conditions~\cite{IEEE802.16CH}. The simulation parameters are presented in Table~\ref{table:sim_param}.  Fig.~\ref{fig:network_topology} illustrates the distribution of wireless users across a $7$-cell network.
\begin{table}[h]
\centering
\begin{tabular}{|l | c|}
\hline
\bf{Parameter} & \bf{Value} \\
\hline
\hline
Number of cells & 7 \\
\hline
Number of Users per cell   &10 \\
\hline
Channel Access  &OFDM/TDD\\
\hline
Carrier Frequency & $1.0$ GHz\\
\hline
Cell Radius  &$500$m \\
\hline
Antenna Model &OMNI - 15 dB\\
\hline
Path Loss Exponent &$3.79$ \\
\hline
Shadowing  &Log-normal  STD = $9$ dB\\
\hline
Channel Coherence Time  &10ms, $\infty$ \\
\hline
Power Update Intervals  &5ms \\
\hline
\end{tabular}
\caption{Simulation parameters.}
\label{table:sim_param}
\end{table}

In the examples, we consider the log-rate as the utility function, i.e.,
$U(\gamma) = \log R(\gamma)$ where $R(\cdot)$ is the achievable rate for a SINR of $\gamma$. This utility function leads to an efficient balance between overall throughput and fairness according to Jain's fairness index (weighted proportional fairness). Assuming $R = \log_2 (1 + \gamma)$, we have $ -1 < \frac{\gamma U''(\gamma)}{U'(\gamma)} < 0$. Then, we have $B = 2$ (see Theorem~\ref{theorem:fixed_point_iterations_min_max_power}) and any $\theta < \theta_{\max} = 1.0$ will guarantee convergence.

\begin{figure}[t]
\centering
\includegraphics[width = 3.5in]{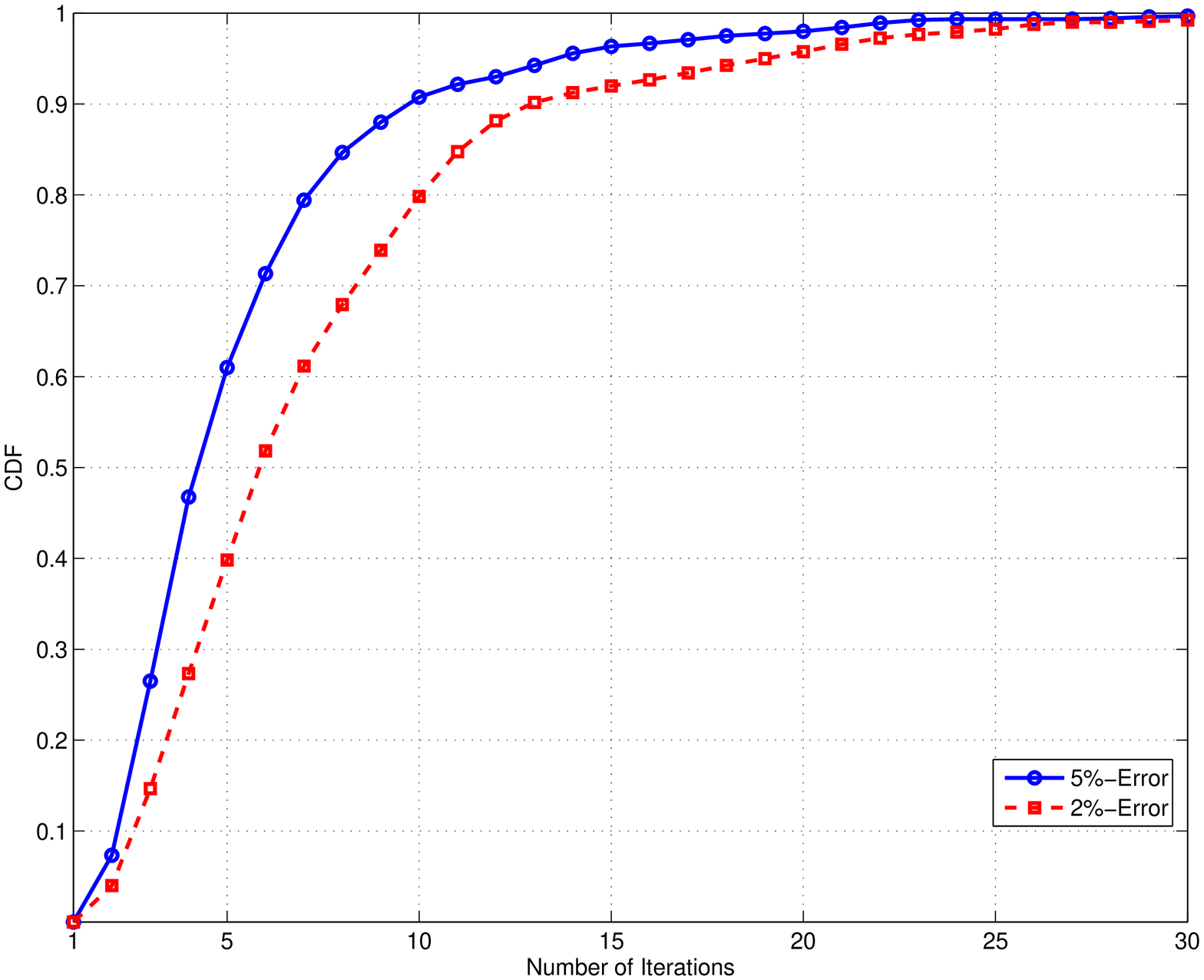}
\caption{CDF of number iterations required to converge to $5\%$ and $1\%$ of optimal solution.}
\label{figure:cdf_iterations}
\end{figure}

\subsection{Rate of Convergence}
\label{sec:numerical_rate_of_convergence}
In our first example, we consider a $7$-cell static wireless network with parameters chosen according to Table~\ref{table:sim_param}. We assume random channel instances and initialize the power control problem with a random starting point every time. We set $R(\gamma) = \log (1 + \frac{\gamma}{5})$ corresponding to a $7$ dB gap to capacity. Fig.~\ref{figure:cdf_iterations} plots the cumulative distribution function (CDF) of the number of iterations required to get to $5\%$ and $2\%$ vicinity of the optimal solution. Based on the results, in $90\%$ of instances, the algorithm converges within 10-15 iterations.

\subsection{Robustness to Channel Variations}
\label{sec:numerical_relay}
Thanks to the fixed-point iteration nature of our proposed algorithm, the convergence speed is at least linear. Therefore, in general, the algorithm can track small changes to the system parameters in a very short time. To this end, we consider the same topology as in the previous example, except that this time we include slow channel fluctuations. Specifically, we assume that the channels are constant within a coherence time of $10$ ms and change independently.

\begin{figure}[t]
\centering
\includegraphics[width = 3.5in]{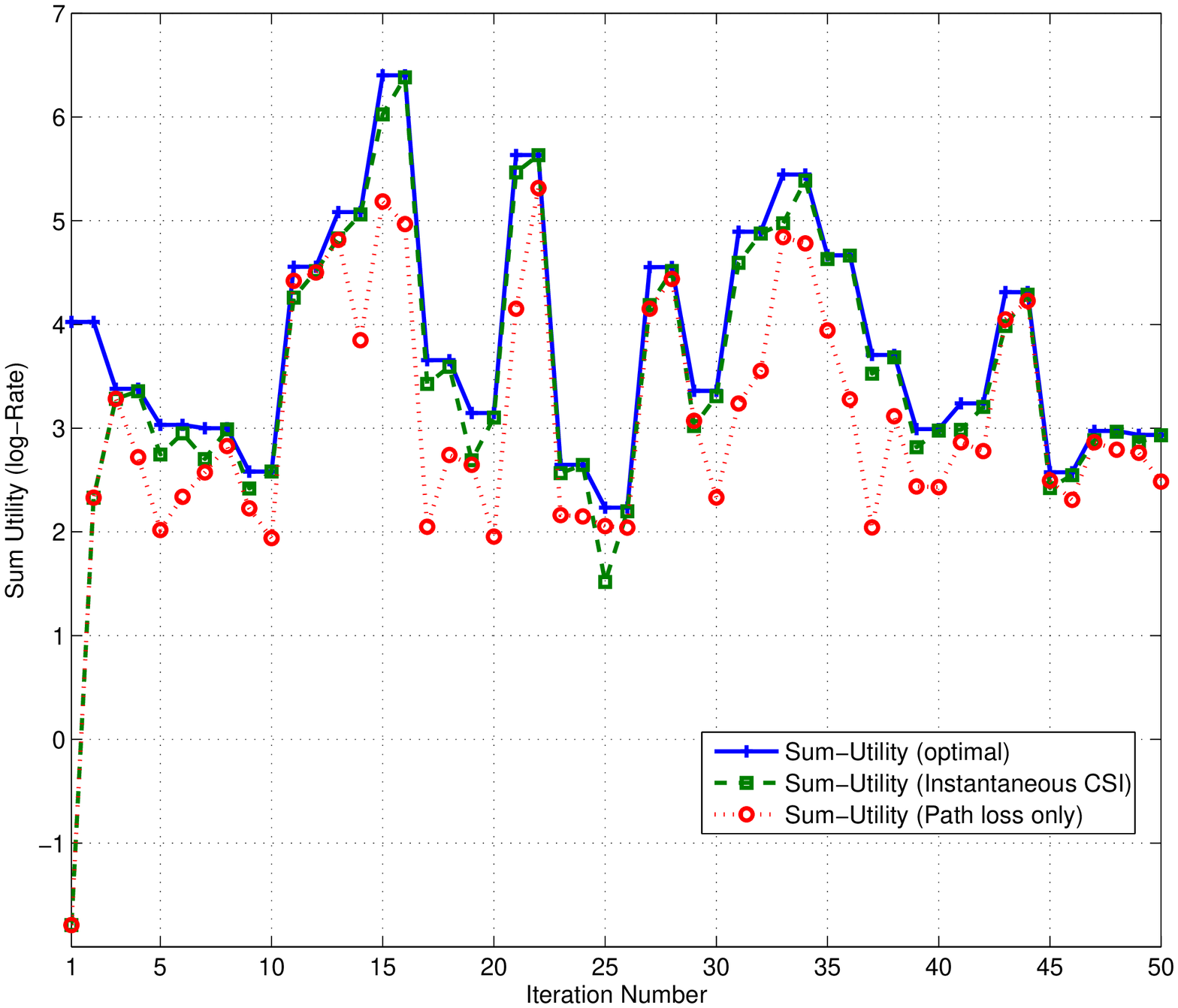}
\caption{Robustness of the power control with respect to small fading channel correlation in between power updates.}
\label{fig:fading_scenario}
\end{figure}

We consider three scenarios: first, the optimal case where the channel powers are updated instantaneously; second where the power update period is half the channel coherence time, and third, where the only channel information available is the average path loss. The results of the simulation are presented in Fig.~\ref{fig:fading_scenario}. As is clear from the figure, the algorithm can efficiently track the channel variations.

\subsection{General Utility Functions}
\label{sec:numerical_general_utility}
In the previous two examples, each node had its own utility function, i.e., there were illustrations of network utility maximization . However, our theoretical development is valid for a broader class of utility functions. In this example we illustrate the ability of the algorithm to optimize power in networks where the utility function is coupled across nodes. We consider a simulation scenario with relay nodes added to our cellular network. We assume in each cell, there is a dedicated relay which helps wireless transmitters with poor channels to the base station, e.g., cell-edge users. The relay operates in decode-and-forward mode~\cite{Laneman:2004}, i.e., the transmission time is divided into two equal time slots where during the first slot the wireless user sends its message to the relay, and during the second time slot, the relay decodes and re-encodes the user's message and forwards forwards it to the base station. It follows that the achievable rate under cooperation is given by
\begin{align}
    R_{c} = \frac{1}{2} \log \left(1 + \frac{\min(\gamma_{wr}, \gamma_{rb})}{\Gamma}\right),
\end{align}
where $\gamma_{wr}$ and $\gamma_{rb}$ are the corresponding SINRs for the wireless transmitter-relay and relay-base station links and $\Gamma \ge 1$ is the decoding loss. The function $\min(\cdot, \cdot)$ has a discontinuous derivative and therefore,  we use the following analytic approximation for the power control algorithm
\begin{align}
\label{eq:min_approx}
\min(x, y) \simeq \frac{-1}{k} \log\left(e^{-kx} + e^{-ky}\right), \; k \gg 1.
\end{align}
Choosing a lager value of $k$ in~\eqref{eq:min_approx} leads to better approximation of $\min(\cdot, \cdot)$ but at the same time increases $B$ defined Theorem~\ref{theorem:fixed_point_iterations_IoT} leading to slower convergence speed (in our simulations, we select $k = 5$ and $\theta = 0.5$ for a $\log$ utility)\footnote{Furthermore, the logarithm of rate based on the proposed approximation is concave over the range $x, y > \frac{1}{k}$.}. Note that the $\min(\cdot,\cdot)$ introduces a coupling of power across the relay and source nodes and the Hessian of the utility function is not diagonal. This is due to the achievable rate for the wireless users using a relay, which is a function two SINRs, namely, user-relay and relay-base station SINRs.

\begin{figure}[t]
\centering
\includegraphics[width = 3.5in]{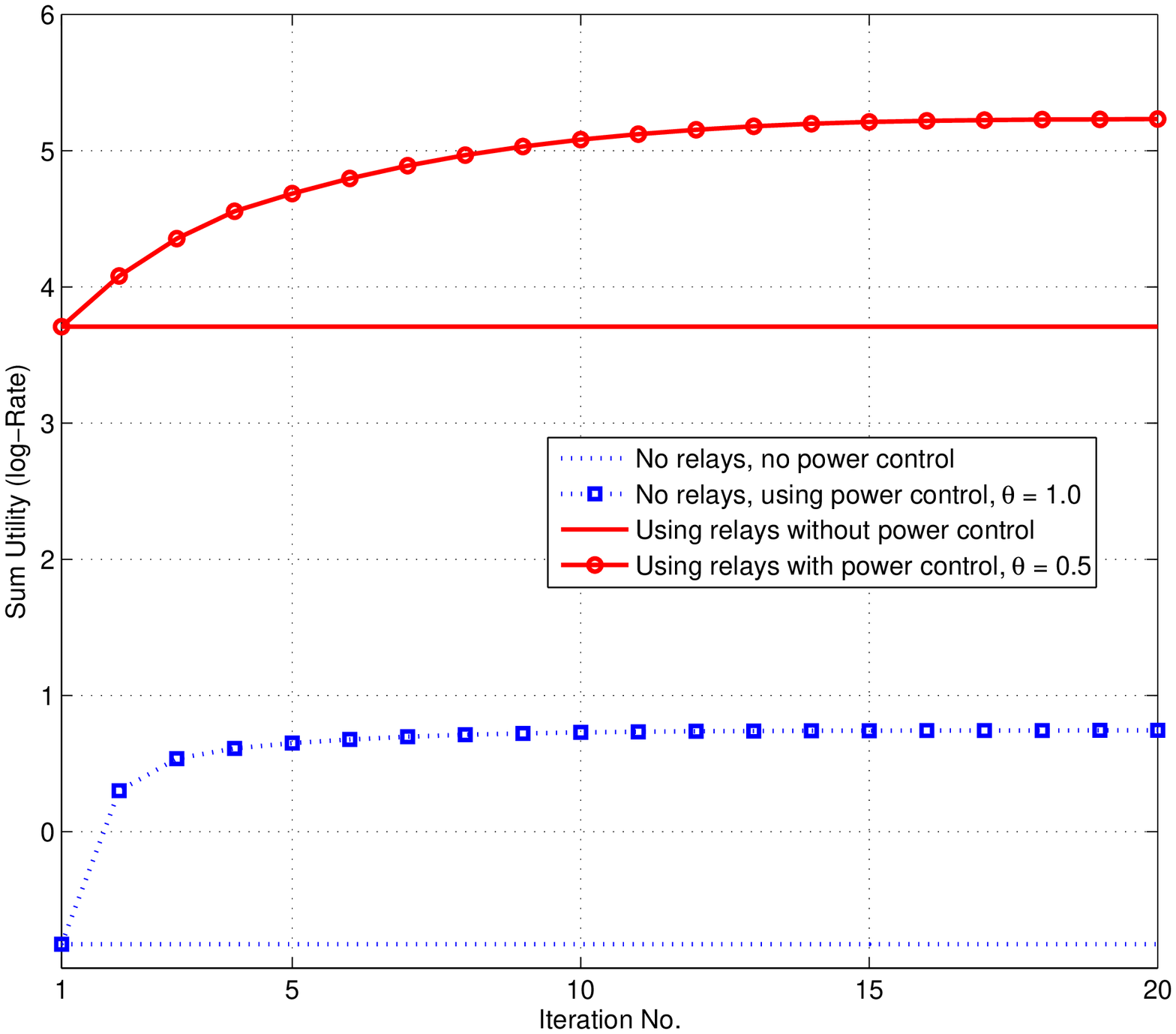}
\caption{Power control in a relay-assisted cellular network.}
\label{fig:relaying_scenario}
\end{figure}

For a wireless transmitter not using the relay, the achievable rate is
\begin{align}
\label{eq:rate_non_coop}
    R_{nc} = \frac{1}{2} \log \left(1 + \frac{\gamma_{sd, 1}}{\Gamma} \right) + \frac{1}{2} \log\left(1 + \frac{\gamma_{sd, 2}}{\Gamma}\right),
\end{align}
where $\gamma_{sd, 1}$ and $\gamma_{sd, 2}$ are the corresponding SINRs in the first and second times slots (the sources of interference are different in the two times slots if relays are used in adjacent cells - the source in the first time slot and the relay in the second). Unfortunately, using~\eqref{eq:rate_non_coop} in the optimization problem~\eqref{eq:optimization_problem_not_convex} leads to a non-convex problem. Therefore, we use the following upper-bound on the achievable rate for the optimization problem
\begin{align}
\label{eq:rate_no_coop_upper_bound}
 R_{nc} \le \log\left(1 + \frac{\gamma_{wb}'}{\Gamma}\right)  \text{ with: } \gamma_{wb}' = \frac{2}{\frac{1}{\gamma_{sd, 1}} + \frac{1}{\gamma_{wb, 2}}}.
\end{align}
In other words, the upper bound in~\eqref{eq:rate_no_coop_upper_bound} is the capacity of the link to the base station when the interference is constant and equal to the average interference received from other cells (from wireless users and relays).
Fig.~\ref{fig:relaying_scenario} illustrates the achieved sum utility in this network for different power control and transmission techniques. In the case of power control, the algorithm is initialized with maximum transmit power and when no power control is used, the wireless transmitters transmit at maximum power. Based on the results in Fig.~\ref{fig:relaying_scenario}, relaying helps with the overall quality of service in cellular networks. Moreover, there is considerable gain in power control after relaying which made feasible through applying the power control algorithm presented in this work. Note that the difference between the convergence speeds of the relayed and non-relayed scenarios is due to the difference in the value of $\theta$. Smaller $\theta$ assures stability at the cost of convergence speed (however, for any $\theta$, the local convergence rate is still linear).

\section{Conclusion}
\label{sec:conclusion}
In this paper, we considered the power control problem in a dense interference environment for a general utility function. We proposed the form of the optimal power vector as the roots of the KKT equations. By introducing an algorithm based on fixed-point iteration, we solved for the power vector and showed that the proposed algorithm converges for a variety of $\log$-concave utility functions. We considered both power constrained and unconstrained scenarios. In particular, we proved that for the scenario with both maximum and minimum power constraints on the wireless transmitters, the proposed algorithm attains the optimal solution for any given $\log$-concave utility function, including functions that do not lead to network utility maximization.

Our numerical results shows that the algorithm converges in a few iterations. Furthermore, due to the fast convergence, the algorithm is very robust to channel variations. Finally, it was shown that the proposed algorithm is capable of solving a general class of utility maximization problems. As an example, power control in a relay-assisted cellular network was demonstrated.

\appendices
\section{Proof of Lemma~\ref{lemma:convexity_of_interference_constraint}}
\label{app:convexity_of_interference_constraint}
The proof shows that the Hessian of $f(\cdot)$ is positive-definite. It readily follows from the definition of $f(\cdot)$ that
\begin{align}
&\bigtriangledown_{\mathbf{x}}^2 f = \left( \bigtriangledown_{\mathbf{x}} f \right)' = \left(\frac{1}{\mathbf{a}^T e^{\mathbf{x}} + a} \mathbf{D}(\mathbf{a}) e^{\mathbf{x}} \right)' = \frac{1}{\mathbf{a}^T e ^ {\mathbf{x}} + a} \times \nonumber \\ &\left(\mathbf{I} - \frac{1}{\mathbf{a}^T e ^ {\mathbf{x}} + a} \mathbf{D}(\mathbf{a}) \mathbf{D}\left(e^{\mathbf{x}} \right) \mathbf{1} \mathbf{1}^T \right) \mathbf{D}(\mathbf{a}) \mathbf{D}\left(e^{\mathbf{x}} \right).
\end{align}
Define $\mathbf{u} = \mathbf{D}(\mathbf{a}) e^{\mathbf{x}}$, i.e., $\mathbf{a}^T e^{\mathbf{x}} = \mathbf{1}^T \mathbf{u}$. Then it follows for the Hessian of $f(\cdot)$ that
\begin{align}
\label{eq:lemma_convexity_int_const_1}
(\mathbf{1}^T \mathbf{u} + a) ^ 2 \bigtriangledown_{\mathbf{x}}^2 f = (\mathbf{1}^T \mathbf{u} + a) \mathbf{D}(\mathbf{u}) - \mathbf{u}\mathbf{u}^T.
\end{align}
 Since $\mathbf{u} > 0$, we can multiplying the sides of~\eqref{eq:lemma_convexity_int_const_1} by have $\mathbf{D}\left(\mathbf{u} \right)^{\frac{1}{2}} \succ 0$ (Note that this conserves the positive semi-definiteness). Then, after multiplication, the right hand side of~\eqref{eq:lemma_convexity_int_const_1} equals
\begin{align}
\label{eq:lemma_convexity_int_const_2}
 (\mathbf{1}^T \mathbf{u} + a) \mathbf{I} - \mathbf{D}\left(\mathbf{u}\right)^{\frac{1}{2}} \mathbf{1} \mathbf{1}^T \mathbf{D}\left(\mathbf{u} \right)^{\frac{1}{2}}.
\end{align}
 Note that the second matrix on the right hand side of~\eqref{eq:lemma_convexity_int_const_2} has a single non-zero eigenvalue of $\parallel \mathbf{D}\left(\mathbf{u} \right)^{\frac{1}{2}} \mathbf{1} \parallel_2 ^ 2 > 0$ and the rest of the eigenvalues are all zero. On the other hand, it simply follows that $\parallel \mathbf{D}\left(\mathbf{u} \right)^{\frac{1}{2}} \mathbf{1} \parallel_2 ^ 2 = \parallel \mathbf{u} \parallel_1 = \mathbf{1}^T \mathbf{u}$. This suggests that the eigenvalues of the matrix on the right hand side of~\eqref{eq:lemma_convexity_int_const_2} are all non-negative and therefore, $f(\cdot)$ is convex. \hfill $\blacksquare$

\section{Proof of Lemma~\ref{lemma:p_alpha_ev_S}}
\label{app:p_alpha_ev_S}
Here we present a short version of the proof. For a more comprehensive study of such matrices we refer the reader to~\cite{Foschini:1993,MChiang:2006} or any text on Perron-Forbenius norms of a positive matrix. Note that for matrix $\mathbf{S}$ we have $s_{ij} > 0 \; \forall i,j$. From Perron-Forbenius theory, $\rho(\mathbf{S})$ corresponds to a positive eigenvalue, namely, $\lambda$. It suffices to show that $\lambda = 1$. Note that if $\lambda > 1$ then there is a vector $\mathbf{p}' > 0$ where $\mathbf{S} \mathbf{p}' = \lambda \mathbf{p'}$. This suggests that $\mathbf{D}(\bm{\gamma}) \mathbf{q}' = \lambda \mathbf{p}'$ where $\mathbf{q'} = \mathbf{V} \mathbf{p}'$ is the interference vector corresponding to the power vector $\mathbf{p}'$.  On the other hand we know that $\mathbf{p}'$ is the eigenvector corresponding to the eigenvalue of $1$ for the matrix $\mathbf{D}(\bm{\gamma}') \mathbf{V}$ where $\gamma' = \frac{\mathbf{p}'}{\mathbf{q}'}$ which implies $\bm{\gamma}' = \lambda \bm{\gamma}$. Therefore,
\begin{align}
\mathbf{D}(\bm{\gamma}') \mathbf{V} \mathbf{p}' = \lambda \mathbf{D}(\bm{\gamma}) \mathbf{V} \mathbf{p}' = \lambda^2 \mathbf{p}'.
\end{align}
This result contradicts with the fact that $\mathbf{p}'$ is the eigenvector corresponding to the eigenvalue of $1$ of the matrix $\mathbf{D}(\bm{\gamma}') \mathbf{V}$. Therefore $\rho[\mathbf{S}] = 1$.
For the second part of the lemma it simply follows from~\eqref{eq:optimal_power_vector} that
\begin{align}
\mathbf{p} = \frac{\mathbf{D}(\bm{\gamma}) \bigtriangledown_{\bm{\gamma}} U}{\mathbf{V}^T \mathbf{D}(\bm{\gamma}) \bigtriangledown_{\bm{\gamma}} U \mathbf{D}\left(\mathbf{q}\right)^{-1}} = \mathbf{p} \frac{\mathbf{D}\left(\mathbf{q}\right)^{-1} \bigtriangledown_{\bm{\gamma}} U}{\mathbf{V}^T \mathbf{D} (\bm{\gamma}) \bigtriangledown_{\bm{\gamma}} U \mathbf{D} \left( \mathbf{q} \right)^{-1}}
\end{align}
where dividing the sides by the non-zero vector $\mathbf{p}$ yields,
\begin{align}
\mathbf{V}^T \mathbf{D}(\bm{\gamma}) \bigtriangledown_{\bm{\gamma}} U \mathbf{D} \left(\mathbf{q}\right) ^{-1}= \mathbf{D}\left( \mathbf{q}\right)^{-1} \bigtriangledown_{\bm{\gamma}} U,
\end{align}
 i.e., $\bigtriangledown_{\bm{\gamma}} U \mathbf{D} \left(\mathbf{q}\right)^{-1}$ is an eigenvector of $\mathbf{S}^T = \mathbf{V}^T \mathbf{D}(\bm{\gamma})$ corresponding to its largest eigenvalue of $1$. By the Perron-Forbenius theorem, this eigenvector is unique and, therefore, it has to be $\bm{\alpha}$.\hfill $\blacksquare$

\section{Proof of Lemma~\ref{lemma:jacobian_phi}}
\label{app:jacobian_phi}
 Consider $\mathbf{p}$ and $\bm{\alpha}$ to be the corresponding right and left eigenvectors of $\mathbf{S}$ at the optimal point. Note that  at the optimal power vector $\mathbf{p}$, we have $\bm{\phi}(\mathbf{p}) = 1$. Then it follows from~\eqref{eq:defn_phi}
\begin{align}
\label{eq:jacobian_phi_step_1}
\bm{\phi}'(\mathbf{p}) &= \left( \frac{\bm{\alpha}}{\mathbf{S}^T \bm{\alpha}} \right)' = \mathbf{D}\left(\left(\mathbf{S}^T \bm{\alpha}\right)^{-1}\right) \bm{\alpha}' \nonumber \\ &- \mathbf{D}\left(\bm{\alpha} \left(\mathbf{S}^T \bm{\alpha}\right)^{-2}\right) \left(\mathbf{S}^T \bm{\alpha}\right)' \nonumber \\
&= \mathbf{D}(\bm{\alpha})^{-1} \left( \bm{\alpha}' - \left(\mathbf{S}^T \bm{\alpha}\right)'\right).
\end{align}
Following $\left(\mathbf{S}^T \bm{\alpha}\right)'$ we have
\begin{align}
\label{eq:jaobian_ST_alpha}
\left(\mathbf{S}^T \bm{\alpha}\right)' &= \left(\mathbf{V}^T \mathbf{D}(\bm{\gamma}) \bm{\alpha} \right)' = \mathbf{V}^T \left(\mathbf{D}(\bm{\gamma}) \bm{\alpha}' + \mathbf{D}(\bm{\alpha}) \bm{\gamma}'\right) \nonumber \\
                                        &= \mathbf{S}^T \bm{\alpha}' + \mathbf{V}^T \mathbf{D}(\bm{\alpha}) \bm{\gamma}'.
\end{align}
For $\bm{\gamma}'$ it follows that
\begin{align}
\label{eq:jacobian_gamma}
\bm{\gamma}' = \left(\frac{\mathbf{p}}{\mathbf{q}}\right)' &= \mathbf{D}(\mathbf{q})^{-1}\mathbf{p}' - \mathbf{D}\left( \frac{\mathbf{p}}{\mathbf{q}^{2}} \right)\mathbf{q}' \nonumber \\
                                                            &= \mathbf{D}(\mathbf{q})^{-1} \left(\mathbf{I} - \mathbf{D}(\bm{\gamma}) \mathbf{V}\right) = \mathbf{D}(\mathbf{q})^{-1} \left(\mathbf{I} - \mathbf{S}\right),
\end{align}
where we use the fact that $\mathbf{q}' = \mathbf{V}$. Finally, for $\bm{\alpha}'$ and from Lemma~\ref{lemma:p_alpha_ev_S} we have
\begin{align}
\label{eq:jacobian_alpha}
\bm{\alpha}' &= \left(\mathbf{D}(\mathbf{q})^{-1} \bigtriangledown_{\bm{\gamma}} U \right)' \nonumber \\
             &= \mathbf{D}(\mathbf{q})^{-1} \frac{d \bigtriangledown_{\bm{\gamma}} U}{d \mathbf{p}} - \mathbf{D}\left(\frac{\bigtriangledown_{\bm{\gamma}} U}{\mathbf{q}^2}  \right) \mathbf{q}' \nonumber \\
            &= \mathbf{D}(\mathbf{q})^{-1} (\bigtriangledown_{\bm{\gamma}}U)' \bm{\gamma}' - \mathbf{D}\left(\frac{\bm{\alpha}}{\mathbf{q}} \right) \mathbf{V}\nonumber \\
            &= \mathbf{D}(\mathbf{q})^{-1} \bigtriangledown_{\bm{\gamma}}^2 U \mathbf{D}(\mathbf{q})^{-1}  \left(\mathbf{I} - \mathbf{S}\right) - \mathbf{D}\left(\frac{\bm{\alpha}}{\mathbf{p}}\right) \mathbf{D}(\bm{\gamma}) \mathbf{V}\nonumber \\
            &= \mathbf{D}(\mathbf{q})^{-1} \bigtriangledown_{\bm{\gamma}}^2 U \mathbf{D}(\mathbf{q})^{-1} \tilde{\mathbf{S}} - \mathbf{D}\left(\frac{\bm{\alpha}}{\mathbf{p}}\right) \mathbf{S}.
\end{align}
Note that in derivations above, $\tilde{\mathbf{S}} = \mathbf{I} - \mathbf{S}$, $\bm{\alpha} = \frac{\bigtriangledown_{\bm{\gamma}} U}{\mathbf{q}}$. Assuming $\mathbf{r} = \frac{\mathbf{p}}{\bm{\alpha}}$, we can use~\eqref{eq:jaobian_ST_alpha},~\eqref{eq:jacobian_gamma},~\eqref{eq:jacobian_alpha} in~\eqref{eq:jacobian_phi_step_1} to obtain
\begin{align}
\label{eq:jacobian_phi_step_3}
\bm{\phi}'(\mathbf{p}) &= \mathbf{D}(\bm{\alpha})^{-1} \left[\tilde{\mathbf{S}}^T \mathbf{D}(\mathbf{q})^{-1} \bigtriangledown_{\bm{\gamma}}^2 U \mathbf{D}(\mathbf{q})^{-1}  \tilde{\mathbf{S}} \right. \nonumber \\
&\left. - \tilde{\mathbf{S}}^T \mathbf{D} \left( \mathbf{r} \right)^{-1} \mathbf{S} - \mathbf{V}^T \mathbf{D} \left( \mathbf{r} \right)^{-1}  \tilde{\mathbf{S}} \right] \nonumber \\
                        &= \mathbf{D}(\bm{\alpha})^{-1} \left[\tilde{\mathbf{S}}^T \mathbf{D}(\mathbf{q})^{-1} \bigtriangledown_{\bm{\gamma}}^2 U \mathbf{D}(\mathbf{q})^{-1} \tilde{\mathbf{S}} \right. \nonumber \\
                        &\left. - \tilde{\mathbf{S}}^T \mathbf{D} \left( \mathbf{r} \right)^{-1} \mathbf{S} - \mathbf{S}^T \mathbf{D} \left( \mathbf{r} \right)^{-1} \tilde{\mathbf{S}} \right].
\end{align}
Finally, by inserting $\mathbf{I} - \tilde{\mathbf{S}}$ in~\eqref{eq:jacobian_phi_step_3} instead of $\mathbf{S}$, and multiplying the sides from left by $\mathbf{D}(\bm{\alpha})$, the expression in Lemma~\eqref{lemma:jacobian_phi} is achieved. \hfill $\blacksquare$

\section{Proof of Lemma~\ref{lemma:sign_of_second_term}}
\label{app:sign_of_second_term}
It follows for the expression in the lemma that that
\begin{align}
\label{eq:sign_of_second_term_derivations}
&\mathbf{D}\left( \mathbf{r} \right)^{\frac{1}{2}} \tilde{\mathbf{S}}^T \mathbf{D}\left( \mathbf{r} \right)^{\frac{-1}{2}} \mathbf{D}\left( \mathbf{r} \right)^{\frac{-1}{2}} \tilde{\mathbf{S}} \mathbf{D}\left( \mathbf{r} \right)^{\frac{1}{2}} \nonumber \\
&- \mathbf{D}\left( \mathbf{r} \right)^{\frac{1}{2}} \tilde{\mathbf{S}}^T \mathbf{D}\left( \mathbf{r} \right)^{\frac{-1}{2}} - \mathbf{D}\left( \mathbf{r} \right)^{\frac{-1}{2}} \tilde{\mathbf{S}} \mathbf{D}\left( \mathbf{r} \right)^{\frac{1}{2}} \nonumber \\
&= \left( \mathbf{D}\left( \mathbf{r} \right)^{\frac{1}{2}} \tilde{\mathbf{S}}^T \mathbf{D}\left( \mathbf{r} \right)^{\frac{-1}{2}} - \mathbf{I} \right) \left(\mathbf{D}\left( \mathbf{r} \right)^{\frac{-1}{2}} \tilde{\mathbf{S}} \mathbf{D}\left( \mathbf{r} \right)^{\frac{1}{2}} - \mathbf{I} \right) - \mathbf{I} \nonumber \\
&= \left( \mathbf{D}\left( \mathbf{r} \right)^{\frac{1}{2}} \mathbf{S}^T \mathbf{D}\left( \mathbf{r} \right)^{\frac{-1}{2}} \right) \left( \mathbf{D}\left( \mathbf{r} \right)^{\frac{-1}{2}} \mathbf{S} \mathbf{D}\left( \mathbf{r} \right)^{\frac{1}{2}} \right) - \mathbf{I}.
\end{align}
Note that all the elements of  the first term in~\eqref{eq:sign_of_second_term_derivations} are positive. From Perron-Forbenius theorem, there is a unique positive eigenvector corresponding to the largest eigenvalue (spectral radius in this case). By observation, we realize that this eigenvector is simply $\left(\bm{\alpha} \mathbf{p}\right)^{\frac{1}{2}}$ which leads to an eigenvalue of $1$. By uniqueness, we realize that~\eqref{eq:sign_of_second_term_derivations} is negative semi-definite with exactly one eigenvalue of $0$ corresponding to the eigenvector $\left( \bm{\alpha} \mathbf{p}\right) ^ {\frac{1}{2}}$.\hfill $\blacksquare$

\section{Proof of Lemma~\ref{lemma:sign_of_jacobian}}
\label{app:sign_of_jacobian}
It follows for the expression in the lemma that
\begin{align}
\label{eq:sign_of_jacobian_rhs_1}
&\mathbf{D}\left( \mathbf{r} \right)^{\frac{1}{2}} \mathbf{D}(\bm{\alpha}) \bm{\phi}'(\mathbf{p}) \mathbf{D}\left(\mathbf{r}\right)^{\frac{1}{2}} =
\mathbf{D}\left( \mathbf{r} \right)^{\frac{1}{2}} \left[ \tilde{\mathbf{S}}^T \left( \mathbf{D}(\mathbf{q})^{-1}\bigtriangledown_{\bm{\gamma}}^2 U \right. \right. \nonumber \\
&\left. \left. \mathbf{D}(\mathbf{q})^{-1} +  \mathbf{D} \left(\mathbf{r} \right)^{-1} \right) \tilde{\mathbf{S}}\right] \mathbf{D}\left( \mathbf{r} \right)^{\frac{1}{2}} + \mathbf{D}\left( \mathbf{r} \right)^{\frac{1}{2}} \left[\tilde{\mathbf{S}}^T \mathbf{D}\left( \mathbf{r}  \right) ^ {-1} \tilde{\mathbf{S}} \right. \nonumber \\
&\left. - \tilde{\mathbf{S}}^T \mathbf{D}\left(\mathbf{r} \right) ^ {-1} - \mathbf{D}\left( \mathbf{r}  \right)^ {-1} \tilde{\mathbf{S}}\right] \mathbf{D}\left( \mathbf{r} \right)^{\frac{1}{2}}.
\end{align}
First note that by Lemma~\ref{lemma:sign_of_second_term}, the second term in~\eqref{eq:sign_of_jacobian_rhs_1} is negative semi-definite with a spectral radius of $2$. On the other hand, the first term depends on the Hessian of the utility function with respect to $\bm{\gamma}$. It follows for the Hessian that
\begin{align}
&\mathbf{D}(\mathbf{q})^{-1} \bigtriangledown_{\bm{\gamma}}^2 U \mathbf{D}(\mathbf{q})^{-1} = \mathbf{D} \left( \frac{\bm{\alpha}}{\mathbf{p}}\right) ^{\frac{1}{2}} \mathbf{D} \left( \frac{\mathbf{p}}{\mathbf{q}} \right)^{\frac{1}{2}} \mathbf{D} \left( \frac{\mathbf{1}}{\bm{\alpha} \mathbf{q}} \right) ^{\frac{1}{2}} \nonumber \\
&\bigtriangledown_{\bm{\gamma}} ^ 2 U \mathbf{D} \left( \frac{\mathbf{p}}{\mathbf{q}} \right)^{\frac{1}{2}} \mathbf{D} \left( \frac{\mathbf{1}}{\bm{\alpha} \mathbf{q}} \right) ^{\frac{1}{2}} \mathbf{D} \left( \frac{\bm{\alpha}}{\mathbf{p}}\right) ^{\frac{1}{2}} \nonumber \\
&=\mathbf{D} \left( \mathbf{r} \right) ^{\frac{-1}{2}} \mathbf{D} \left(\frac{\bm{\gamma}}{\bigtriangledown_{\bm{\gamma}} U} \right) ^{\frac{1}{2}} \bigtriangledown_{\bm{\gamma}}^2 U \mathbf{D} \left(\frac{\bm{\gamma}}{\bigtriangledown_{\bm{\gamma}} U} \right) ^ {\frac{1}{2}} \mathbf{D}\left( \mathbf{r} \right) ^{\frac{-1}{2}} \nonumber \\
&= \mathbf{D} \left( \mathbf{r} \right) ^{\frac{-1}{2}} \mathbf{M}_U \mathbf{D} \left( \mathbf{r} \right) ^{\frac{-1}{2}},
\end{align}
where $\mathbf{M}_U := \mathbf{D} \left(\frac{\bm{\gamma}}{\bigtriangledown_{\bm{\gamma}} U} \right) ^{\frac{1}{2}} \bigtriangledown_{\bm{\gamma}}^2 U \mathbf{D} \left(\frac{\bm{\gamma}}{\bigtriangledown_{\bm{\gamma}} U} \right) ^ {\frac{1}{2}}$. On the other hand, from~\eqref{eq:log_concavity} and the assumption in~\eqref{eq:log_concavity_bound}, we have
\begin{align}
\label{eq:hessian_bound}
-(B - 1) \mathbf{I} \preceq \mathbf{M}_U + \mathbf{I}   \preceq 0.
\end{align}
By using~\eqref{eq:hessian_bound} in the first terms of~\eqref{eq:sign_of_jacobian_rhs_1} it follows that
\begin{align}
&\mathbf{D} \left( \mathbf{r} \right) ^ {\frac{1}{2}} \left[ \tilde{\mathbf{S}}^T \left( \mathbf{D}(\mathbf{q} )^ {-1}\bigtriangledown_{\bm{\gamma}}^2 U \mathbf{D}(\mathbf{q} )^ {-1} + \mathbf{D} \left( \mathbf{r} \right)  ^ {-1} \right) \tilde{\mathbf{S}}\right] \mathbf{D} \left( \mathbf{r} \right) ^ {\frac{1}{2}} \nonumber \\
\label{eq:sign_first_term_jacobian}
&= \mathbf{D} \left( \mathbf{r} \right) ^ {\frac{1}{2}} \tilde{\mathbf{S}}^T \mathbf{D}(\mathbf{r})^{\frac{-1}{2}} \left(\mathbf{M}_U + \mathbf{I} \right) \mathbf{D}(\mathbf{r})^{\frac{-1}{2}} \tilde{\mathbf{S}} \mathbf{D} \left( \mathbf{r} \right) ^ {\frac{1}{2}}  \\
\label{eq:spectral_radius_first_term_jacobian_1}
& \succeq -(B - 1) \mathbf{D} \left( \mathbf{r} \right) ^{\frac{1}{2}} \tilde{\mathbf{S}}^T \mathbf{D} \left(\mathbf{r} \right)^ {-1}  \tilde{\mathbf{S}} \mathbf{D} \left( \mathbf{r} \right) ^{\frac{1}{2}}.
\end{align}
From~\eqref{eq:hessian_bound} and~\eqref{eq:sign_first_term_jacobian}, we see that the first term in~\eqref{eq:sign_of_jacobian_rhs_1} is negative semi-definite. Furthermore, following~\eqref{eq:spectral_radius_first_term_jacobian_1}, we realize that
\begin{align}
\label{eq:norm_tilde_ST_tilde_S}
\rho\left[ \mathbf{D} \left( \mathbf{r} \right) ^{\frac{1}{2}} \tilde{\mathbf{S}}^T \mathbf{D} \left( \mathbf{r}  \right) ^{-1} \tilde{\mathbf{S}} \mathbf{D} \left( \mathbf{r} \right) ^{\frac{1}{2}} \right] = \rho \left( \tilde{\mathbf{S}}^T \tilde{\mathbf{S}} \right) = \parallel \tilde{\mathbf{S}} \parallel_2^2,
\end{align}
where~\eqref{eq:norm_tilde_ST_tilde_S} comes from the properties of matrix norms\footnote{For any positive semi-definite matrix $\mathbf{A}$, we have $\rho(\mathbf{A}^T \mathbf{A}) = \parallel \mathbf{A} \parallel_2^2$.}. It follows that
\begin{align}
\parallel \tilde{\mathbf{S}} \parallel_2 &= \max_{\parallel \mathbf{x} \parallel_2 = 1} \parallel \tilde{\mathbf{S}} \mathbf{x} \parallel_2 =\max_{\parallel \mathbf{x} \parallel_2 = 1} \parallel \mathbf{x} - \mathbf{S}\mathbf{x}\parallel_2 \nonumber \\
\label{eq:norm_of_tilde_ST_tilde_final}
& \le \max_{\parallel \mathbf{x} \parallel_2 = 1} \parallel \mathbf{x} \parallel_2 + \parallel \mathbf{S} \mathbf{x} \parallel_2 \nonumber \\
&\le \max_{\parallel \mathbf{x} \parallel_2 = 1} \parallel \mathbf{x} \parallel_2  + \rho(\mathbf{S}) \parallel \mathbf{x} \parallel_2 < 2,
\end{align}
where the last result is an strict inequality since $\mathbf{x}$ cannot be $\mathbf{p}$ (in such a case, $\parallel \tilde{\mathbf{S}} \mathbf{p} \parallel = 0$).
Finally, from~\eqref{eq:sign_of_jacobian_rhs_1},~\eqref{eq:spectral_radius_first_term_jacobian_1} and~\eqref{eq:norm_of_tilde_ST_tilde_final} we realize that $\mathbf{D}\left( \mathbf{r} \right)^{\frac{1}{2}} \mathbf{D}(\bm{\alpha}) \bm{\phi}'(\mathbf{p}) \mathbf{D}\left(\mathbf{r}\right)^{\frac{1}{2}}$ is negative semi-definite and also
\begin{align}
\rho(\mathbf{D} (\mathbf{p}) \bm{\phi}'(\mathbf{p})) \le 4(B - 1) + 2.
\end{align}
\hfill $\blacksquare$

\section{Proof of Corollary~\ref{corollary:finite_norm_fixed_point}}
\label{app:finite_norm_fixed_point}
First note that $\mathbf{p}$ is a fixed point of~\eqref{eq:finite_norm_fixed_point}. More importantly, for any $r > 0$, $r \mathbf{p}$ is~\emph{not} a fixed point of~\eqref{eq:finite_norm_fixed_point}. From the Jacobian of~\eqref{eq:finite_norm_fixed_point} it follows that
\begin{align}
\label{eq:jacobian_fininte_norm_phi}
\left(\bm{\phi}_{\theta}\left(\frac{\mathbf{p}}{\parallel \mathbf{p} \parallel_2}\right) \right)' &= \bm{\phi}_{\theta}'\left( \frac{\mathbf{p}}{\parallel \mathbf{p} \parallel_2} \right) \left( \frac{\mathbf{p}}{\parallel \mathbf{p} \parallel_2} \right)' \nonumber \\
&= \left( \mathbf{I} + \theta \mathbf{D}\left( \mathbf{r} \right) \mathbf{A} \right) \left(\mathbf{I} - \mathbf{p} \mathbf{p}^T \right),
\end{align}
where $\mathbf{A} = \mathbf{D}(\bm{\alpha}) \bm{\phi}'(\mathbf{p})$ is a symmetric negative semi-definite matrix with a single eigenvalue of $0$ corresponding to the eigenvector $\mathbf{p}$ (See Lemma~\ref{app:jacobian_phi}). From linear algebra we know that the eigenvectors of $\mathbf{J}$ are linearly independent and form a basis in $\mathbb{R}^N$. Consider $\mathbf{v}_1$ to $\mathbf{v}_N$ as the eigenvectors of $\mathbf{A}$ corresponding to the eigenvalues $\lambda_1$ to $\lambda_N$ where $\lambda_1 = 0$, and $\lambda_i \le \lambda_{i - 1}$ are all real and negative. Furthermore, $\lambda_N > -4B + 2$ where $B$ is defined in~\eqref{theorem:fixed_point_iterations_IoT}.

We need to show that the spectral radius of~\eqref{eq:jacobian_fininte_norm_phi} is less than $1$. It follows that
\begin{align}
\rho\left[\left(\bm{\phi}_{\theta}\left(\frac{\mathbf{p}}{\parallel \mathbf{p} \parallel_2}\right) \right)' \right] &= \rho\left[ \left(\mathbf{I} + \theta \mathbf{D}\left( \mathbf{r} \right) \mathbf{A} \right) \left(\mathbf{I} - \mathbf{p} \mathbf{p}^T \right) \right] \nonumber \\
&= \rho\left[ \left(\mathbf{I} + \hat{\mathbf{A}} \right) \mathbf{Q} \right],
\end{align}
where
\begin{align}
    \hat{\mathbf{A}} = \mathbf{D} \left( \mathbf{r} \right) ^ {\frac{1}{2}} \mathbf{A} \mathbf{D} \left( \mathbf{r} \right) ^ {\frac{-1}{2}}, \\
    \mathbf{Q} = \mathbf{I} - \mathbf{D} \left( \mathbf{r} \right) ^ {\frac{1}{2}} \mathbf{p} \mathbf{p}^T \mathbf{D} \left( \mathbf{r} \right) ^ {\frac{-1}{2}}.
\end{align}
Note that $\mathbf{A}$ and $\hat{\mathbf{A}}$ share the same eigenvalues. Furthermore, $\mathbf{D}\left(\mathbf{r} ^ {-1} \right)^{\frac{1}{2}} \mathbf{v}_i$ is the eigenvectors of $\hat{\mathbf{J}}$ corresponding to the eigenvalue $\lambda_i$. By assumption, we have $\mathbf{v}_1 = \mathbf{p}$ and since $\mathbf{Q} \mathbf{D}\left(\mathbf{r} ^ {-1} \right)^{\frac{1}{2}} \mathbf{p} = 0$. For the other eigenvectors $\mathbf{v}_i$, $ 1 < i \le N$ it yields
\begin{align}
\left(\mathbf{I} + \hat{\mathbf{A}}\right) \mathbf{Q} \mathbf{D}\left( \mathbf{r}\right)^{\frac{-1}{2}} \mathbf{v}_i &= \left(\mathbf{I} + \hat{\mathbf{A}}\right) \mathbf{D}\left( \mathbf{r} \right)^{\frac{-1}{2}} \mathbf{v}_i \nonumber \\
&= (1 + \theta \lambda_i) \mathbf{D}\left( \mathbf{r} \right)^{\frac{-1}{2}} \mathbf{v}_i,
\end{align}
which readily suggests
\begin{align}
\rho\left[\left(\bm{\phi}_{\theta}\left(\frac{\mathbf{p}}{\parallel \mathbf{p} \parallel_2}\right) \right)' \right] = \max_{i, 2 \le i \le N} \mid 1 + \theta \lambda_i \mid < 1. \nonumber \hfill \blacksquare
\end{align}

\section{Proof of Lemma~\ref{lemma:jacobian_phi_non_IoT}}
\label{app:jacobian_phi_non_IoT}
First note that for $\bm{\phi}'(\mathbf{p})$ we have
\begin{align}
\label{eq:jacobian_phi_non_IoT_step_2}
&\bm{\phi}'(\mathbf{p}) = \mathbf{D} \left( \frac{\mathbf{1}}{\mathbf{S}^T \bm{\alpha}} \right) \bm{\alpha}' - \mathbf{D} \left( \frac{\bm{\alpha}}{\mathbf{S}^T \bm{\alpha}}\right) \mathbf{D}\left(\frac{\mathbf{1}}{\mathbf{S}^T \bm{\alpha}} \right) \left( \mathbf{S}^T \bm{\alpha}\right)' \nonumber \\
                    &= \mathbf{D}\left( \frac{\bm{\phi}(\mathbf{p})}{\bm{\alpha}} \right) \left( \bm{\alpha}'  - \left( \mathbf{I} + \mathbf{D}(\bm{\phi}(\mathbf{p})) - \mathbf{I} \right) \left(\mathbf{S}^T \bm{\alpha} \right)' \right) \nonumber \\
                    &= \mathbf{D}(\bm{\phi}(\mathbf{p})) \left[ \bm{\phi}_{\bm{\zeta} = 0}'(\mathbf{p}) - \left(\mathbf{D}(\bm{\phi}(\mathbf{p})) - \mathbf{I} \right) \left( \mathbf{S}^T \bm{\alpha}\right)' \right],
\end{align}
where $\bm{\phi}_{\bm{\zeta} = 0}'(\mathbf{p})$ is the Jacobian of $\bm{\phi}(\mathbf{p})$ in the interference limited case evaluated at $\mathbf{p}$. For $\bm{\phi}_{\theta}(\mathbf{p})'$ it follows that
\begin{align}
&\left(\bm{\phi}_{\theta}(\mathbf{p})\right)' = \theta \mathbf{D}(\bm{\phi}(\mathbf{p})) + \theta \mathbf{D}(\mathbf{p}) \bm{\phi}'(\mathbf{p})  + (1 - \theta) \mathbf{I} \nonumber  \\
\label{eq:jacobian_phi_non_IoT_final}
&= \mathbf{D}(\bm{\phi}(\mathbf{p})) \left( \mathbf{I} + \theta \mathbf{D}(\mathbf{p}) \bm{\phi}_{\bm{\zeta} = 0}'(\mathbf{p}) \right) -\left( \mathbf{D}(\bm{\phi}(\mathbf{p})) - \mathbf{I} \right) \times \nonumber \\
&\left( (1 - \theta) \mathbf{I} + \theta \mathbf{D}\left(\frac{\mathbf{p} \bm{\phi}(\mathbf{p})}{\bm{\alpha}}\right) \left(\mathbf{S}^T \bm{\alpha}\right)' \right).
\end{align}
Note that $\bm{\phi}_{\theta, \bm{\zeta} = \mathbf{0}}(\mathbf{p}) = \mathbf{I} + \theta \mathbf{D}(\mathbf{p}) \bm{\phi}_{\bm{\zeta} = 0}'(\mathbf{p})$ is simply the Jacobian of $\bm{\phi}_{\theta}(\mathbf{p})$ assuming $\bm{\zeta} = 0$. The only difference is that, since $\mathbf{p}$ is no more an eigenvector of $\bm{\phi}_{\theta, \bm{\zeta} = 0}(\mathbf{p})$ corresponding to an eigenvalue of $1$, $\rho \left[ \bm{\phi}_{\theta, \bm{\zeta} = 0}(\mathbf{p}) \right]$ is~\emph{strictly} less than $1$. The second terms in~\eqref{eq:jacobian_phi_non_IoT_final} is also a bounded matrix. Also note that $\left(\mathbf{D}(\bm{\phi}(\mathbf{p})) - I\right)_{i,i} = 0$ for all $ i > n$. \hfill $\blacksquare$
\bibliographystyle{IEEETran}
\bibliography{pcreflist}

\end{document}